\documentstyle[eqsecnum,aps]{revtex}

\newcommand{\begit}{\begin{itemize}}
\newcommand{\enit}{\end{itemize}}
\newcommand{\beq}{\begin{equation}} 
\newcommand{\eeq}{\end{equation}} 
\newcommand{\beqa}{\begin{eqnarray}} 
\newcommand{\eeqa}{\end{eqnarray}} 
\newcommand{\p}{\partial}    
\newcommand{\pr}{^\prime}    
\begin{document}
\draft
\preprint{noidea}

\title{Mu and Tau Neutrino Thermalization and Production\\ in Supernovae: 
Processes and Timescales}
\author{Todd A. Thompson}
\address{Department of Physics, \\
The University of Arizona, Tucson 85721 \\
thomp@physics.arizona.edu}
\author{Adam Burrows}
\address{Steward Observatory and Department of Astronomy,\\
The University of Arizona, Tucson 85721 \\
burrows@jupiter.as.arizona.edu}
\author{Jorge E. Horvath}
\address{Instituto Astron\'{o}mico e Geof\'{i}sico, \\
Universidade de S\~{a}o Paulo, Brazil \\
foton@orion.iagusp.usp.br}
\date{\today}
\maketitle 
\begin{abstract}
We investigate the rates of production and thermalization of $\nu_\mu$ and 
$\nu_\tau$ neutrinos at temperatures and densities relevant to core-collapse 
supernovae and protoneutron stars.  Included are contributions from electron 
scattering, electron-positron annihilation, nucleon-nucleon bremsstrahlung, 
and nucleon scattering. For the scattering processes, in order to incorporate 
the full scattering kinematics at arbitrary degeneracy, the structure function 
formalism developed by Reddy et al. (1998) and Burrows and Sawyer (1998) is 
employed.  Furthermore, we derive formulae for the total and differential 
rates of nucleon-nucleon bremsstrahlung for arbitrary nucleon degeneracy in 
asymmetric matter. We find that electron scattering dominates nucleon 
scattering as a thermalization process at low neutrino energies 
($\varepsilon_\nu\lesssim 10$ MeV), but that nucleon scattering is always 
faster than or comparable to electron scattering above 
$\varepsilon_\nu\simeq10$ MeV.  In addition, for $\rho\gtrsim 10^{13}$ 
g cm$^{-3}$, $T\lesssim14$ MeV, and neutrino energies $\lesssim60$ MeV, 
nucleon-nucleon bremsstrahlung always dominates electron-positron annihilation 
as a production mechanism for $\nu_\mu$ and $\nu_\tau$ neutrinos.
\end{abstract}

\setlength{\baselineskip}{12pt}

\section{Introduction}
\label{sec:intro}

The cores of protoneutron stars and core-collapse supernovae are 
characterized by mass densities of order $\sim 10^{10}-10^{14}$ g cm$^{-3}$  
and temperatures that range from $\sim 1$ to $50$ MeV.  
The matter is composed predominantly of nucleons, electrons, positrons, and 
neutrinos of all species.  For $\nu_\mu$ and $\nu_\tau$ types 
(collectively `$\nu_\mu$s'), which carry away 50$-$60\% of the 
$\sim 2-3\times 10^{53}$ ergs liberated during collapse and explosion, 
the prevailing opacity and production processes are $\nu_\mu$-electron 
scattering, $\nu_\mu$-nucleon scattering, electron-positron annihilation 
($e^+e^-\leftrightarrow \nu_\mu\bar{\nu}_\mu$), and nucleon-nucleon 
bremsstrahlung. While all of these processes contribute for the electron 
types ($\nu_e$s and $\bar{\nu}_e$s), the charged-current absorption  
processes $\nu_e n\leftrightarrow p e^-$ and $\bar{\nu}_ep\leftrightarrow ne^+$ 
dominate their opacity so completely that in this paper
we address only $\nu_\mu$ production and thermalization.  

Supernova theorists had long held \cite{lamb_pethick} that $\nu_\mu$-nucleon 
scattering was unimportant as a mechanism for neutrino equilibration.
While this process was included as a source of 
opacity \cite{bruenn_1985,bhf_1995}, it served only to redistribute the 
neutrinos in space, not in energy.  In contrast, $\nu_\mu$-electron scattering 
was thought to dominate $\nu_\mu$ neutrino thermalization.  
In addition, the only $\nu_\mu\bar{\nu}_\mu$ pair production mechanisms 
employed in full supernova calculations were 
$e^+e^-\leftrightarrow \nu_\mu\bar{\nu}_\mu$ and plasmon decay 
($\gamma_{pl}\leftrightarrow \nu_\mu\bar{\nu}_\mu$) \cite{bruenn_1985}; 
nucleon-nucleon bremsstrahlung was neglected as a source.  Recent developments, 
however, call both these practices into question and motivate a re-evaluation of 
these opacities in the supernova context.
Analytic formulae have recently been derived 
\cite{reddy_1998,burrows_sawyer,bs_1999} which 
include the full nucleon kinematics and 
Pauli blocking in the final state at arbitrary nucleon degeneracy.  
These efforts reveal that the average rate of energy transfer
in $\nu_\mu$-nucleon scattering may surpass previous estimates by an order of 
magnitude 
\cite{burrows_sawyer,hannestad,keil_1995,janka_1996,raffelt_seckel,sigl_1997}. 
Hence, this process may compete with $\nu_\mu$-electron scattering as an 
equilibration mechanism.  Similarly, estimates for the total nucleon-nucleon 
bremsstrahlung rate have been obtained  
\cite{hannestad,burrows_1999,friman,fsb_1975} which indicate that this process 
might compete with  $e^+e^-$ annihilation.

These results suggest that the time is ripe for a technical study
of the relative importance of each process for production or thermalization.
To conduct such a study, we consider $\nu_\mu$ neutrinos in an isotropic 
homogeneous thermal bath of scatterers and absorbers.  In this system, the 
full transport problem is reduced to an evolution of the  neutrino 
distribution function (${\cal F}_\nu$) in energy space alone.   Although 
this is a simplification of the true problem, it provides a
theoretical laboratory in which to analyze the rates both for equilibration 
of an initial neutrino
distribution function with dense nuclear matter and for production of the 
neutrinos themselves. 
From these rates we determine the importance and particular character of 
each process, and discover
in which energy, temperature, or density regime each dominates.
We employ a general prescription for solving the Boltzmann equation in 
this system with the full energy redistribution collision term.  We compare 
quantitatively, via direct numerical evolution of an arbitrary neutrino 
distribution function,
the rates for thermalization and production by each process, at all neutrino
energies.  Furthermore, we present the total nucleon-nucleon bremsstrahlung 
rate for arbitrary nucleon degeneracy and 
derive the single $\nu_\mu$ and $\bar{\nu}_\mu$ production spectra.  This 
facilitates a more comprehensive
evaluation of its relative importance in neutrino production than has 
previously been possible.

In \S \ref{sec:genform}, we discuss the general form
of the Boltzmann equation and our use of it to study $\nu_\mu$ equilibration 
and production rates.  In \S \ref{sec:interactions}, we provide
formulae for each of the four processes we consider: $\nu_\mu$-nucleon 
scattering, $\nu_\mu$-electron scattering, and $\nu_\mu\bar{\nu}_\mu$ pair 
production via both nucleon-nucleon bremsstrahlung
and $e^+e^-$ annihilation. In \S \ref{sec:results}, we present the results of 
our equilibration calculations, showing the time evolution of $\nu_\mu$ 
distribution functions as influenced by each of these processes individually.  
We include plots of thermalization and production rates for each process as a 
function of neutrino energy and time.  For the scattering interactions we 
include figures of the time evolution of the net energy transfer to the medium 
as a function of incident neutrino energy.
We repeat this analysis at points in temperature, density, and composition 
space relevant to supernovae and protoneutron stars, taken from snapshots of 
a stellar profile during a realistic collapse calculation \cite{bhf_1995}.  
Using these results, we discuss the relative importance of each process in 
shaping the emergent $\nu_\mu$ spectrum. In \S \ref{sec:conclusions}, we
recapitulate our findings and conclusions.

\section{The Boltzmann Equation}
\label{sec:genform}
The static (velocity$\,=\,0$) Boltzmann equation for the evolution of the 
neutrino distribution function (${\cal F}_\nu$),  including Pauli blocking 
in the final state, and for a spherical geometry, is
\beq
\left(\frac{1}{c}\frac{\p}{\p t}+\mu\frac{\p}{\p r}+
\frac{1-\mu^2}{r}\frac{\p}{\p \mu}\right)
{\cal F}_\nu=(1-{\cal F}_\nu)j_\nu-{\cal F}_\nu\chi_\nu,
\label{genboltz}
\eeq
where $t$ is the time, $r$ is the radial coordinate, and $\mu(=\cos\theta)$ 
is the cosine of the zenith angle.
$j_\nu$ and $\chi_\nu$ are the total $source$ and $sink$, respectively. 
For emission and absorption, 
$j_\nu$ is the emissivity and $\chi_\nu$ is the extinction coefficient. 
For scattering, both $j_\nu$ 
and $\chi_\nu$ are energy redistribution integrals which couple one neutrino 
energy bin with all the others. 
The matrix element and associated phase-space integrations which comprise 
$j_\nu$ and $\chi_\nu$ for 
electron and nucleon scattering yield the probability that a given collision 
will scatter 
a particle into any angle or energy bin.  A full transport calculation couples 
energy and angular bins to each other through the right hand side of 
eq. (\ref{genboltz}). 

In a homogeneous, isotropic thermal bath of scatterers and absorbers no 
spatial or angular gradients exist.  Consequently, the Boltzmann equation 
becomes 
\beq
\frac{1}{c}\frac{\p{\cal F}_\nu}{\p t}=
(1-{\cal F}_\nu)j_\nu-{\cal F}_\nu \chi_\nu.
\label{boltzmann1}
\eeq
By dealing with this system, the transport problem reduces to an evolution 
of ${\cal F}_\nu$ in just energy and time. 
Note that for scattering processes, both $j_\nu$ and $\chi_\nu$
require an integral over the scattered neutrino distribution function 
${\cal F}_\nu\pr$.  Similarly, in evolving ${\cal F}_\nu$ via the production 
and absorption processes, $j_\nu$ and $\chi_\nu$ involve an integration over 
the anti-neutrino distribution function ${\cal F}_{\bar{\nu}}$.  Therefore, 
${\cal F}_{\bar{\nu}}$ must be evolved simultaneously with ${\cal F}_\nu$. 
While $j_\nu$ and $\chi_\nu$  may be fairly complicated integrals 
over phase-space, the numerical solution of eq. (\ref{boltzmann1}) is 
straightforward.  

Given an arbitrary initial ${\cal F}_\nu$, 
we divide the relevant energy range into $n$ energy bins.  
We then solve eq. (\ref{boltzmann1}) for each bin individually and 
explicitly. Angular integrals over scattering cosines which appear in the 
$\nu_\mu$-nucleon and $\nu_\mu$-electron scattering 
formalism, as well as the electron energy integration needed for $e^+e^-$ 
annihilation, are carried out with a 4-point Gauss-Legendre integration scheme.  
The double integral over dimensionless nucleon momentum variables needed to 
obtain the contribution from nucleon-nucleon bremsstrahlung is computed using 
nested 16-point Gauss-Laguerre quadratures.

\subsection{Rates for ${\cal F}_\nu$ Evolution and Energy Transfer}

Scattering, emission, and absorption processes, at a given neutrino energy 
($\varepsilon_\nu$), produce and remove neutrinos from the phase-space density 
at that energy.  The former achieves this by transferring energy
to the matter during scattering, the latter two by emitting or absorbing 
directly from that bin.  The Boltzmann equation can then be written in terms of
an {\it in} and an {\it out} channel, the former a {\it source} and the latter 
a {\it sink}:
\beq
\frac{\p{\cal F}_\nu}{\p t}=\left.\frac{\p{\cal F}_\nu}{\p t}\right|_{in}-
\left.\frac{\p{\cal F}_\nu}{\p t}\right|_{out}.
\label{boltztime}
\eeq
Consequently, for any interaction, there are two rates to consider: 
the rate for scattering or production into a given energy bin ($\Gamma_{in}$) 
and the inverse rate for scattering or absorption out of that bin 
($\Gamma_{out}$).  The rates $cj_\nu$ and $c\chi_\nu$ yield timescales
for an interaction to occur, but fail, in the case of the former, to fold 
in Pauli blocking in the final state.  Equation (\ref{boltztime}) includes 
these effects and provides natural timescales for ${\cal F}_\nu$ evolution:
\beq
\Gamma_{in}\,=\,\frac{1}{{\cal F}_\nu}
\left.\frac{\p{\cal F}_\nu}{\p t}\right|_{in}\,=\,
\frac{(1-{\cal F}_\nu)}{{\cal F}_\nu}\,cj_\nu
\label{gammain}
\eeq
and
\beq
\Gamma_{out}\,=\,\frac{1}{{\cal F}_\nu}
\left.\frac{\p{\cal F}_\nu}{\p t}\right|_{out}\,=\,
c\chi_\nu.
\label{gammaout}
\eeq
Note that although eq. (\ref{gammaout}) does not explicitly contain a 
Pauli blocking term, $\chi_\nu$ contains an integral over 
$(1-{\cal F}_\nu\pr)$, in the case of scattering, and 
an appropriate final-state blocking term, in the case of absorption.  
At a given $\varepsilon_\nu$, then, $\Gamma_{in}$ incorporates information
about the $\nu_\mu$ phase-space density at that energy.  Conversely, at that 
same $\varepsilon_\nu$, $\Gamma_{out}$ contains information about the 
phase-space at all other energies.  Regardless of the initial distribution, 
$\p{\cal F}_\nu/\p t=0$ in equilibrium.  This implies 
$\Gamma_{in}=\Gamma_{out}$ in equilibrium and, hence, we build in a test 
for the degree to which the system has thermalized.

Just as there are distinct rates for the $in$ and $out$ channels of the 
Boltzmann equation during equilibration, so too are there distinct scattering 
energy transfers.  For $\nu_\mu$ scattering with a scatterer $s$ 
(electron or nucleon), at a specific $\varepsilon_\nu$, two thermal average 
energy transfers can be defined; 
\beq
\langle\omega\rangle_{in}\,=
\,\int d^3p_\nu\pr\,\omega\,{\cal F}_\nu\pr\,
{\cal I}^{in}\left[\nu_\mu s\leftarrow \nu_\mu\pr s\pr\right]
\,/\,\int d^3p_\nu\pr\,{\cal F}_\nu\pr\,
{\cal I}^{in}\left[\nu_\mu s\leftarrow \nu_\mu\pr s\pr\right]
\label{win}
\eeq
and
\beq
\langle\omega\rangle_{out}\,=
\,\int d^3p_\nu\pr\,\omega\,(1-{\cal F}_\nu\pr)\,
{\cal I}^{out}\left[\nu_\mu s\rightarrow \nu_\mu\pr s\pr\right]
\,/\,\int d^3p_\nu\pr\,(1-{\cal F}_\nu\pr)\,
{\cal I}^{out}\left[\nu_\mu s\rightarrow \nu_\mu\pr s\pr\right],
\label{wout}
\eeq
where primes denote the scattered neutrino, 
$\omega(=\varepsilon_\nu-\varepsilon_\nu\pr)$ is the 
energy transfer, and ${\cal I}^{in}$ and ${\cal I}^{out}$ are the kernels 
for scattering into and out of a given energy bin, respectively.  As a 
consequence of detailed balance between the {\it in} and {\it out} channels 
of the Boltzmann equation, ${\cal I}^{in}=e^{-\beta\omega}{\cal I}^{out}$, 
where $\beta=1/k_B T$ and $T$ is the matter temperature.  
(The scattering kernels are discussed in detail in \S \ref{sec:interactions} 
for both scattering processes.)  Note that the denominators in 
eqs. (\ref{win}) and (\ref{wout}), up to constants which divide out in the 
definitions of $\langle\omega\rangle_{in}$ and $\langle\omega\rangle_{out}$, 
are just $j_\nu$ and $\chi_\nu$, respectively.

In an effort to provide more than one measure of the timescale for 
${\cal F}_\nu$ equilibration due to scattering and to make contact with 
previous neutrino thermalization studies 
\cite{tubbs_scatter,tubbs_thermal,tubbs_equil} 
we also define a set of timescales in terms of $\langle\omega\rangle_{out}$ 
and the higher $\omega$-moment, $\langle\omega^2\rangle_{out}$;
\beq
\Gamma_D=
c\chi_\nu\,\left|\frac{\langle\omega\rangle_{out}}{\varepsilon_\nu}\right|
\label{gammad}
\eeq
and
\beq
\Gamma_E=c\chi_\nu\,\frac{\langle\omega^2\rangle_{out}}{\varepsilon_\nu^2}.
\label{gammae}
\eeq
$\Gamma_D$ is the rate for shifting the centroid of a given distribution and
$\Gamma_E$ is the rate for spreading an initial distribution 
\cite{tubbs_scatter}. In contrast with the work of 
\cite{tubbs_scatter,tubbs_thermal,tubbs_equil}, we include the full effects 
of Pauli blocking in the final state, allowing us to deal consistently with 
cases in which the $\nu_\mu$s are partially degenerate.

\section{Individual Interactions}
\label{sec:interactions}

This section details the source and sink terms necessary to solve the 
Boltzmann equation for the time-evolution of ${\cal F}_\nu$. Sections 
\S\ref{subsec:nscatt} and \S\ref{subsec:escatt} are dedicated to the 
presentation and discussion of the collision terms for $\nu_\mu$-nucleon 
and $\nu_\mu$-electron scattering, respectively.  Section \S\ref{subsec:pairpr}
describes the Legendre series expansion approximation and the use of it to 
compute the contribution to the  Boltzmann equation, the pair emissivity, 
and the single $\nu_\mu$ spectrum due to 
$e^+e^-\leftrightarrow\nu_\mu\bar{\nu}_\mu$.  
Our derivations of $j_\nu$ and $\chi_\nu$, as well as the single and pair 
spectra from nucleon-nucleon bremsstrahlung at arbitrary nucleon degeneracy 
and in the non-degenerate limit, are presented in \S\ref{subsec:bremss}.
In what follows, we take 
$G^2\simeq1.55\times10^{-33}$ cm$^{3}$ MeV$^{-2}$ s$^{-1}$,
$\sin^2\theta_W\simeq0.231$, and employ natural units in which $\hbar=c=k_B=1$.

\subsection{Nucleon Scattering: 
$\nu_\mu n \leftrightarrow \nu_\mu n$ and $\nu_\mu p \leftrightarrow \nu_\mu p$}
\label{subsec:nscatt}

Researchers working on supernova and protoneutron star evolution have recently 
re-evaluated the issue of energy transfer via $\nu_\mu$-nucleon scattering 
\cite{burrows_sawyer,hannestad,keil_1995,janka_1996,raffelt_seckel}.  
Originally, the assumption was made that the nucleons were stationary 
\cite{lamb_pethick}.  If a neutron of mass $m_n$  is at rest with respect to 
an incoming neutrino of energy $\varepsilon_\nu$, one finds that the energy 
transfer ($\bar{\omega}$) is $\sim-\varepsilon^2_\nu/m_n$.  
For $\varepsilon_\nu\,=\,10$ MeV, $\bar{\omega}\sim-0.1$ MeV, a 
fractional energy lost of 1\%.  Using these simple kinematic arguments 
and disregarding neutrino and nucleon Pauli blocking, one finds that the
thermalization rate for $\nu_\mu$-electron scattering should be approximately 
a factor of 20 larger than that for $\nu_\mu$-nucleon scattering.
In the context of interest, however, at temperatures of order 10 MeV and 
mass densities of order $10^{13}$ g cm$^{-3}$, free nucleons are not 
stationary, but have thermal velocities.  The fractional energy exchange 
per collision, in the case of $\nu_\mu$-neutron scattering, 
is then $\sim p_n/m_nc$ \cite{burrows_sawyer}. For $T\sim10$ MeV this gives 
a $\sim$10$-$20\% change in $\varepsilon_\nu$ per collision.  This calls the 
naive estimate of the relative importance of $\nu_\mu$-nucleon scattering as 
a thermalization process into question and a more complete exploration of the 
relative importance of the two scattering processes is necessary. 

Recently, analytic formulae have been derived which include the full 
kinematics of $\nu_\mu$-nucleon scattering at arbitrary nucleon degeneracy  
\cite{reddy_1998,burrows_sawyer,bs_1999}.  At the temperatures and 
densities encountered in the supernova context non-interacting nucleons are 
not relativistic.  Due to nucleon-nucleon interactions, however,  at and 
around nuclear density ($\sim 2.68\times 10^{14}$ g cm$^{-3}$), the nucleon's 
effective mass drops and is expected to be comparable with its Fermi momentum 
\cite{reddy_1998}.  In such a circumstance, a relativistic description of the 
$\nu_\mu$-nucleon scattering interaction is warranted.
In addition, spin and density correlation effects engendered by these 
nucleon-nucleon interactions have been found to suppress the $\nu_\mu$-nucleon 
interaction rate by as much as a factor of $\sim2-3$ 
\cite{burrows_sawyer,raffelt_seckel,sigl_1997}.

In this study, we focus on $\nu_\mu$ equilibration rates at densities 
$\lesssim 1\times 10^{14}$ g cm$^{-3}$ where it is still unclear if 
nucleon-nucleon interactions will play an important role. This ambiguity 
is due in part to uncertainties both in the nuclear equation of state 
and the nucleon-nucleon interaction itself.  For this reason we choose 
to treat the nucleons as non-relativistic and non-interacting, thereby 
ignoring collective effects which might enhance or reduce the 
$\nu_\mu$-nucleon scattering rate.  Making these assumptions, 
we find that $j_\nu$ and $\chi_\nu$ in eq. (\ref{boltzmann1}) are given by
\beq
j_\nu=\frac{G^2}{(2\pi)^3}\int d^3\vec{p}^{\,\prime}_\nu
\,{\cal I}_{{\rm NC}}\,{\cal F}_\nu\pr \,e^{-\beta\omega}
\label{nscattj}
\eeq
and
\beq
\chi_\nu=\frac{G^2}{(2\pi)^3}\int d^3\vec{p}^{\,\prime}_\nu
\,{\cal I}_{{\rm NC}}\,[1-{\cal F}_\nu\pr],
\label{nscattx}
\eeq
where $\beta=1/T$, $\vec{p}^{\,\prime}_\nu$ is the final state neutrino 
momentum, and $\omega$ is the energy transfer.  In eqs. (\ref{nscattj}) 
and (\ref{nscattx}), the neutral-current scattering kernel is given by 
\beq
{\cal I}_{{\rm NC}}=S(q,\omega)\,[(1+\mu)V^2+(3-\mu)A^2],
\label{Inc}
\eeq
where $\mu(=\cos\theta)$ is the cosine of the scattering angle between 
incident and final state neutrinos and $S(q,\omega)$ is the dynamic structure 
function.  In eq. (\ref{Inc}), $V$ and $A$ are the appropriate vector and 
axial-vector coupling constants; for $\nu_\mu$-neutron scattering, 
$V=-1/2$ and $A=-1.26/2$.  The dynamic structure function is
\beqa
S(q,\omega)&=&2\int\frac{d^3\vec{p}}{(2\pi)^3}\,{\cal F}\,
(1-{\cal F}\pr)\,(2\pi)\delta(\omega+\varepsilon-\varepsilon\pr) \nonumber \\
&=&2\,{\rm Im}\Pi^{(0)}(q,\omega)\,(1-e^{-\beta\omega})^{-1}
\label{sqw}
\eeqa
where $q=|p_\nu-p_\nu\pr|=[\varepsilon_\nu^2+\varepsilon_\nu^{\prime\,2}-
2\varepsilon_\nu\varepsilon_\nu\pr\mu]^{1/2}$ is the magnitude of the momentum 
transfer, and ${\cal F}$ and ${\cal F}\pr$ are the incident and scattered 
nucleon distribution functions, respectively.  In eq. (\ref{sqw}), $\vec{p}$ 
is the incident nucleon momentum, $\varepsilon$ is the incident nucleon energy, 
and $\varepsilon\pr$  is the scattered nucleon energy. 
The imaginary part of the free polarization is given by 
\cite{burrows_sawyer,fetter_walecka}
\beq
{\rm Im}\Pi^{(0)}(q,\omega)=\frac{m^2}{2\pi\beta q}{\rm ln}
\left[\frac{1+e^{-Q^2+\eta}}{1+e^{-Q^2+\eta-\beta\omega}}\right],
\label{npolarization}
\eeq
where 
\beq
Q=\left(\frac{m\beta}{2}\right)^{1/2}
\left(-\frac{\omega}{q}+\frac{q}{2m}\right),
\eeq
$\eta$ is the nucleon degeneracy ($\mu/T$), and $m$ is the nucleon mass.  
The factor $e^{-\beta\omega}$ which appears in eq. (\ref{nscattj}) is a 
consequence of the fact that $S(q,-\omega)=e^{-\beta\omega}S(q,\omega)$, 
itself a consequence of detailed balance between the $in$ and $out$ channels 
of the Boltzmann equation. The dynamic structure function can be thought of 
as a correlation function which connects $\varepsilon_\nu$ and 
$\varepsilon_\nu\pr$. 

The $\phi$ angular integrations implicit in eqs. (\ref{nscattj}) and 
(\ref{nscattx}) can be computed trivially assuming the isotropy of 
${\cal F}_\nu$.  Furthermore, defining the coordinate system with the momentum 
vector of the incident neutrino, the scattering angle and the direction cosine 
are equivalent.  Combining these two equations in the Boltzmann equation for 
the evolution of ${\cal F}_\nu$ due to neutral-current $\nu_\mu$-nucleon 
scattering, we obtain 
\beq
\frac{\p{\cal F}_\nu}{\p t}=\frac{G^2}{(2\pi)^2}\int_0^\infty 
d\varepsilon_\nu\pr\varepsilon_\nu^{\prime\,2}\int_{-1}^{1}\,d\mu
\,\,{\cal I}_{{\rm NC}}
\left\{[1-{\cal F}_\nu]\,{\cal F}_\nu\pr e^{-\beta\omega}-
{\cal F}_\nu[1-{\cal F}_\nu\pr]\right\}.
\label{workingnscatt}
\eeq

\subsection{Electron Scattering: 
$\nu_\mu e^-\leftrightarrow \nu_\mu e^-$} 
\label{subsec:escatt}

At the temperatures and densities encountered in supernovae and 
protoneutron stars, electrons are highly relativistic.  A formalism 
analogous to that used for $\nu_\mu$-nucleon scattering is desired in 
order to include the full electron kinematics at arbitrary electron 
degeneracy.  Reddy et al. \cite{reddy_1998} have developed a relativistic 
generalization of the structure function formalism described
in \S\ref{subsec:nscatt}.  They obtain a set of polarization functions which 
characterize the relativistic medium's response to a neutrino probe in terms 
of polylogarithmic functions.  In analogy with eq. (\ref{workingnscatt}), 
we can write the Boltzmann equation for the evolution
of ${\cal F}_\nu$ due to $\nu_\mu$-electron scattering, as
\beq
\frac{\p{\cal F}_\nu}{\p t}=
\frac{G^2}{(4\pi)^3}
\int d^3p_\nu\pr\,
{\cal I}^{\,r}_{{\rm NC}}
\left\{[1-{\cal F}_\nu]\,{\cal F}_\nu\pr e^{-\beta\omega}-
{\cal F}_\nu[1-{\cal F}_\nu\pr]\right\},
\label{escatt}
\eeq
where ${\cal I}^{\,r}_{{\rm NC}}$ is the relativistic neutral-current 
scattering kernel for $\nu_\mu$s, analogous to ${\cal I}_{{\rm NC}}$ in 
eq. (\ref{Inc}).  All the physics of the interaction is contained in 
${\cal I}^{\,r}_{NC}$, which can be written as
\beq
{\cal I}^{\,r}_{{\rm NC}}=
{\rm Im}\{\Lambda^{\alpha\beta}\Pi^R_{\alpha\beta}\}(1-e^{-\beta\omega})^{-1}.
\eeq
As in the non-relativistic case, ${\cal I}^{\,r}_{{\rm NC}}$ is composed of 
the lepton tensor, 
\beq
\Lambda^{\alpha\beta}=8[2k^\alpha k^\beta+(k\cdot q)g^{\alpha\beta}-
(k^\alpha q^\beta+q^\alpha k^\beta)- 
i\epsilon^{\alpha\beta\mu\nu}k^\mu q^\nu],
\eeq
which is just the squared and spin-summed matrix element for the
scattering process written in terms of $k_\alpha$, the incident $\nu_\mu$ 
four-momentum, and $q_\alpha\left(=(\omega,\vec{q}\,)\right)$, the 
four-momentum transfer.  The scattering kernel also contains 
the retarded polarization tensor, $\Pi^R_{\alpha\beta}$, which is
directly analogous to the free polarization in the non-relativistic case 
given in eq. (\ref{sqw}).  The retarded polarization tensor is related to 
the causal polarization by
\beq
{\rm Im}\,\Pi^R_{\alpha\beta}=
\tanh\left(-\frac{1}{2}\beta\omega\right){\rm Im}\,\Pi_{\alpha\beta}
\eeq
and
\beq
\Pi_{\alpha\beta}=
-i\int\frac{d^4p}{(2\pi)^4}{\rm Tr}[G_{e}(p)J_\alpha G_{e}\pr(p+q)J_\beta].
\label{causalpol}
\eeq
In eq. (\ref{causalpol}), $p_\alpha$ is the electron four-momentum and 
$J_\alpha$ is the current operator.  The electron Green's functions 
($G_e$ and $G_e\pr$), explicit in the free polarization, connect points 
in electron energy space and characterize the effect of the interaction 
on relativistic electrons.  The polarization tensor can be written in 
terms of a vector part, an axial-vector part, and a mixed part, so that
\beq
\Pi_{\alpha\beta}=V^2\Pi^V_{\alpha\beta}+A^2\Pi^A_{\alpha\beta}-
2V\hspace{-2pt}A\,\Pi^{VA}_{\alpha\beta}.
\eeq
In turn, the vector part of the polarization tensor can be written in terms 
of two independent components, $\Pi_T$ and $\Pi_L$.  In contrast with 
eq. (\ref{Inc}), since $v/c\sim 1$ for the electrons, the angular terms
which were dropped from the matrix element in the non-relativistic case, 
leading to a single structure function, must now be retained.  
${\cal I}^{\,r}_{{\rm NC}}$ can then be written as a set of three structure 
functions \cite{reddy_1998}:
\beq
{\cal I}^{\,r}_{{\rm NC}}=8[A{\cal S}_1(q,\omega)+
{\cal S}_2(q,\omega)+B{\cal S}_3(q,\omega)](1-e^{-\beta\omega})^{-1},
\label{Irnc}
\eeq
where $A=(4\varepsilon_\nu\varepsilon_\nu\pr+q_\alpha^2)/2q^2$ and 
$B=\varepsilon_\nu+\varepsilon_\nu\pr$.  These structure functions can 
be written in terms of the vector parts of the retarded polarization 
tensor ($\Pi^R_T$ and $\Pi^R_L$), the axial part ($\Pi^R_A$), and the 
mixed part ($\Pi^R_{VA}$):
\beq
{\cal S}_1(q,\omega)\,=
\,(V^2+A^2)\,\,\left[\,{\rm Im}\Pi_L^R(q,\omega)+
{\rm Im}\Pi_T^R(q,\omega)\,\right],
\label{s1}
\eeq
\beq
{\cal S}_2(q,\omega)\,=
\,(V^2+A^2)\,\,{\rm Im}\Pi_T^R(q,\omega)-
A^2{\rm Im}\Pi_A^R(q,\omega),
\label{s2}
\eeq
and
\beq   
{\cal S}_3(q,\omega)\,=
\,2V\hspace{-.06cm}A\,\,{\rm Im}\Pi_{VA}^R(q,\omega).
\label{s3}
\label{s1s2s3}
\eeq
The retarded polarization functions, in terms of differences 
between polylogarithmic integrals, can be found in 
Appendix \ref{app:escatt}. 

\subsection{Electron-Positron Annihilation: 
$e^+e^-\leftrightarrow \nu_\mu\bar{\nu}_\mu$}
\label{subsec:pairpr}

Fermi's Golden Rule for the total volumetric emission 
rate for the production of $\nu_\mu$s via electron-positron 
annihilation can be written as 
\beq
Q=\int\frac{d^3\vec{p}}{(2\pi)^3 2\varepsilon}
\frac{d^3\vec{p}^{\,\prime}}{(2\pi)^3 2\varepsilon\pr}
\,\frac{d^3\vec{q}_\nu}{(2\pi)^32\varepsilon_\nu}
\frac{d^3\vec{q}_{\bar{\nu}}}{(2\pi)^3 2\varepsilon_{\bar{\nu}}}
\,\,\varepsilon_\nu\,\left(\frac{1}{4}\sum_s|{\cal M}^2|\right)
\,(2\pi)^4\delta^4({\bf {\rm{P}}}) \,
\Xi[{\cal F}],
\label{totpairpr}
\eeq
where
\beq
\Xi[{\cal F}]=
(1-{\cal F}_\nu)(1-{\cal F}_{\bar{\nu}}){\cal F}_{e^-}{\cal F}_{e^+},
\label{blocking}
\eeq
and $\delta^4({\bf {\rm{P}}})$ conserves four-momentum. 
In eq. (\ref{totpairpr}), $p_\alpha(=(\varepsilon,\vec{p}\,))$
and $p\pr_\alpha(=(\varepsilon\pr,\vec{p}^{\,\prime}))$ are the 
four-momenta of the electron and positron, respectively,
and $q_\nu^\alpha(=(\varepsilon_\nu,\vec{q_\nu}))$ and 
$q_{\bar{\nu}}^{\alpha}(=(\varepsilon_{\bar{\nu}},\vec{q}_{\bar{\nu}}))$
are the four-momenta of the $\nu_\mu$ and $\bar{\nu}_\mu$, respectively. 
The process of electron-positron annihilation into a 
neutrino/anti-neutrino pairs is related to neutrino-electron 
scattering considered in \S\ref{subsec:escatt} via a 
crossing symmetry.  In order to make the problem tractable, 
we follow the standard procedure \cite{bruenn_1985} of
expanding the production kernel in a Legendre series in the 
scattering angle to first order (see Appendix \ref{app:pairpr}).  
Near the neutrinospheres, at densities which render neutrino transport 
diffusive this approximation holds. In a full neutrino transport 
algorithm, however, which must handle both the diffusion and 
free-streaming limits, the second-order term, with proper closure 
relations, must be used in the semi-transparent regime between the 
neutrinospheres and the shock \cite{pons_98}.  Having made this 
approximation, including only the zeroth- and first-order terms, 
the single $\nu_\mu$ spectrum is 
\beq
\frac{dQ}{d\varepsilon_\nu}=
(1-{\cal F}_\nu)\,\frac{\varepsilon_\nu^3}{8\pi^4}\,\int_0^\infty\,
d\varepsilon_{\bar{\nu}}\,\varepsilon_{\bar{\nu}}^2
\,\,\Phi_0^p(\varepsilon_\nu,\varepsilon_{\bar{\nu}})
\,(1-{\cal F}_{\bar{\nu}}),
\label{spectrum}
\eeq
where $\Phi_0^p(\varepsilon_\nu,\varepsilon_{\bar{\nu}})$ 
is the zeroth-order production kernel expansion coefficient,
an integral over the electron energy 
(see Appendix \ref{app:pairpr}) \cite{bruenn_1985}.
With the differential spectrum or emissivity ($dQ/d\varepsilon_\nu$) 
in hand, it is a simple matter to extract the contribution to the 
Boltzmann equation due to $e^+e^-$ annihilation.  As eq. (\ref{spectrum})
already contains the $\nu_\mu$ blocking factor, the contribution to the
Boltzmann equation, the $in$ channel explicit in eq. (\ref{boltztime}), 
can be written as \cite{bruenn_1985}
\beq
\left.\frac{\p {\cal F}_\nu}{\p t}\right|_{in}=
\frac{1}{4\pi}\frac{(2\pi)^3}{\varepsilon_\nu^3}
\frac{dQ}{d\varepsilon_\nu}.
\label{j_spec}
\eeq
In order to obtain the $out$ channel for absorption due to 
$e^+e^-$ annihilation, we need only replace 
${\cal {F}}_{e^-}{\cal {F}}_{e^+}$ in eq. (\ref{blocking}) 
with an electron/positron blocking term, 
$(1-{\cal {F}}_{e^-})(1-{\cal {F}}_{e^+})$, and replace the $\nu_\mu$ 
and $\bar{\nu}_\mu$ blocking terms
in eq. (\ref{spectrum}) with ${\cal {F}}_{\nu}{\cal {F}}_{\bar{\nu}}$. 
Finally, the Boltzmann equation for the evolution of ${\cal F}_\nu$ in 
time due to $e^+e^-\leftrightarrow\nu_\mu\bar{\nu}_\mu$ can be written as
\beq
\frac{\p {\cal F}_\nu}{\p t}=
\frac{2G^2}{(2\pi)^3}\int_0^\infty\hspace{-.2cm}
d\varepsilon_{\bar{\nu}}\,\varepsilon_{\bar{\nu}}^{2}\,\int_0^{\epsilon}
\hspace{-.1cm}d\varepsilon 
\,H_o(\varepsilon_\nu,\varepsilon_{\bar{\nu}},\varepsilon)\,
\left\{(1-{\cal F}_\nu)
(1-{\cal F}_{\bar{\nu}}){\cal F}_{e^-}{\cal F}_{e^+}-
{\cal F}_\nu{\cal F}_{\bar{\nu}}
(1-{\cal F}_{e^-})(1-{\cal F}_{e^+})\right\},
\label{workingpairpr}
\eeq
where $\epsilon=\varepsilon_\nu+\varepsilon_{\bar{\nu}}$ and 
$H_o(\varepsilon_\nu,\varepsilon_{\bar{\nu}},\varepsilon)$ is 
given in eq. (\ref{polynomials}).  In solving eq. (\ref{workingpairpr}), 
${\cal F}_{\bar{\nu}}$ must be evolved simultaneously with ${\cal F}_\nu$.
To do so, in addition to making the appropriate changes to the vector and 
axial-vector coupling constants, $V$ and $A$, one needs to integrate 
over $\varepsilon_\nu$ instead of $\varepsilon_{\bar{\nu}}$. 
Note that the electron and positron distribution functions appear 
explicitly in eq. (\ref{workingpairpr}). We take these distributions 
to be Fermi-Dirac at temperature $T$ and with 
$\eta_e$ determined by $T$, $\rho$, and $Y_e$.

Equation (\ref{totpairpr}) may also be used to find the total 
volumetric $\nu_\mu\bar{\nu}_\mu$ pair spectrum by replacing 
$\varepsilon_\nu$ in the numerator with $\epsilon$.  Ignoring neutrino 
blocking in the final state one can show that \cite{dicus}
\beq
Q_{\nu_\mu\bar{\nu}_\mu}\,\simeq\,
2.09\times10^{24}\left(\frac{T}{{\rm MeV}}\right)^9f(\eta_e)
\,\,\,\,\,\,{\rm ergs\,\,\,cm^{-3}\,\,\,s^{-1}},
\label{eetspec}
\eeq
where
\beq
f(\eta_e)\,=
\,\frac{F_4(\eta_e)F_3(-\eta_e)+
F_4(-\eta_e)F_3(\eta_e)}{2F_4(0)F_3(0)},
\eeq
and 
\beq
F_n(y)=\int_0^\infty\frac{x^n}{e^{x-y}+1}\,dx
\eeq
are the Fermi integrals. 

\subsection{Nucleon-Nucleon Bremsstrahlung}
\label{subsec:bremss}

The importance of nucleon-nucleon bremsstrahlung in late-time neutron 
star cooling has been acknowledged for some time \cite{friman,fsb_1975}.  
Recently, however, this process has received 
more attention as a contributor of $\nu_\mu\bar{\nu}_\mu$ pairs and as 
an energy transport mechanism in both
core-collapse supernova and nascent neutron star evolution 
\cite{hannestad,burrows_1999,brinkmann,suzuki_93}.  The contribution from
nucleon-nucleon bremsstrahlung is a composite of neutron-neutron ($nn$), 
proton-proton ($pp$), and neutron-proton ($np$) bremsstrahlung.  
Fermi's Golden Rule for the total volumetric emissivity of single $\nu_\mu$s 
due to $nn$, $pp$, or $np$ bremsstrahlung, including $\nu_\mu$ and 
$\bar{\nu}_\mu$ blocking in the final state, is given by
\beq
Q=\int\left[\,\prod^4_{i=1}\frac{d^3\vec{p}_i}{(2\pi)^3}\right]
\,\frac{d^3\vec{q}_\nu}{(2\pi)^32\varepsilon_\nu}
\frac{d^3\vec{q}_{\bar{\nu}}}{(2\pi)^3 2\varepsilon_{\bar{\nu}}}
\,\varepsilon_\nu\,\left(s\sum|{\cal M}|^2\right)
\,(2\pi)^4\delta^4({\bf {\rm{P}}}) \,
\Xi[{\cal F}]
\label{totbrem1}
\eeq
where
\beq
\Xi[{\cal F}]=
{\cal F}_1{\cal F}_2(1-{\cal F}_3)(1-{\cal F}_4)
(1-{\cal F}_\nu)(1-{\cal F}_{\bar{\nu}}).
\eeq
The product of differential phase space factors in 
eq. (\ref{totbrem1}) includes a term for each of the four nucleons 
involved in the process; 1 and 2 denote initial-state nucleons whereas 
3 and 4 denote final-state nucleons. In eq. (\ref{totbrem1}), $s$ is a 
symmetry factor for identical initial-state fermions,
$\vec{q}_\nu$ is the neutrino three-momentum, $\varepsilon_\nu$ is the 
neutrino energy, and the four-momentum conserving delta function is 
explicit.  In a one-pion exchange model for the nucleon-nucleon 
interaction, the spin-summed matrix element can be approximated 
by \cite{friman,brinkmann} 
\beq
\sum_s|{\cal M}|^2\simeq 64G^2g_A^2\left(\frac{f}{m_\pi}\right)^4
\left[\left(\frac{k^2}{k^2+m^2_\pi}\right)^2+\dots\right]
\epsilon^{-2}
(\varepsilon_\nu\varepsilon_{\bar{\nu}}-
\vec{q}_\nu\cdot\hat{k}\,\vec{q}_{\bar{\nu}}\cdot\hat{k})
\eeq
where $\epsilon=\varepsilon_\nu+\varepsilon_{\bar{\nu}}$, $k$ is the 
magnitude of the nucleon momentum transfer, $g_A\simeq-1.26$, $f\sim 1$ 
is the pion-nucleon coupling, and $m_\pi$ is the mass of the pion.  
In order to make the 18-dimensional phase-space integration in 
eq. (\ref{totbrem1}) tractable we assume the quantity in square 
brackets to be of order unity, but possibly as low as 0.1 
\cite{burrows_1999}.  To acknowledge our ignorance, we introduce the 
factor, $\zeta$, and assume these momentum terms are constant.  
Furthermore, we neglect the momentum of the neutrinos relative to the 
momentum of the nucleons.  We are left with a simple, but general, form 
for the bremsstrahlung matrix element:
\beq
\sum|{\cal M}|^2\simeq
\,A\zeta\frac{\varepsilon_\nu\varepsilon_{\bar{\nu}}}{\epsilon^2},
\label{simpleM}
\eeq
where $A=64G^2g_A^2f^4/m_\pi^4$.
In the case of $nn$ or $pp$ bremsstrahlung, as appropriate for identical 
particles in the initial state, the symmetry factor ($s$) in 
eq. (\ref{totbrem1}) is $1/4$.  Such a symmetry factor does not enter 
for the mixed-nucleon process, $np$, which is still further enhanced
by the fact that a charged pion mediates the nucleon exchange 
\cite{brinkmann}.  
This increases the matrix element in eq. (\ref{simpleM}) 
by a factor of $7/3$ in the degenerate nucleon limit and $\sim 5/2$ in 
the non-degenerate limit \cite{brinkmann}.  
Considering the already substantial simplifications made by choosing not 
to handle the momentum terms directly, we will adopt the more conservative 
$4\times(7/3)$ enhancement for the $np$ matrix element.
The total volumetric emission rate combining all processes is just 
$Q_{tot}=Q_{nn}+Q_{pp}+Q_{np}$.
What remains is to reduce eq. (\ref{totbrem1}) to a useful expression 
in asymmetric matter and at arbitrary neutron and proton degeneracy.

Following ref. \cite{brinkmann}, we define new momenta, 
$p_{\pm}=(p_1\pm p_2)/2$ and $p_{3c,4c}=p_{3,4}-p_+$, 
new direction cosines, $\gamma_1=p_+\cdot p_-/|p_+||p_-|$ and 
$\gamma_c=p_+\cdot p_{3c}/|p_+||p_{3c}|$, and let $u_i=p_i^2/2mT$. 
Furthermore, we note that $d^3p_1d^3p_2=8d^3p_+d^3p_-$. Using the 
three-momentum conserving delta function, we can do the $d^3\vec{p}_4$ 
integral trivially.  Rewriting eq. (\ref{totbrem1}) with these definitions,
we find that
\beq
Q=2As\zeta(2mT)^{9/2}(2\pi)^{-9}\int d\varepsilon_\nu\,\varepsilon_\nu^3
\int d\varepsilon_{\bar{\nu}}\,du_-\,du_+\,du_{3c}\,d\gamma_1\,d\gamma_c\,
(\varepsilon_{\bar{\nu}}/\epsilon)^2\,(u_-u_+u_{3c})^{1/2}\,\delta(E)
\,\,\Xi[{\cal F}],
\label{bremQ1}
\eeq
where 
\beq
\delta(E)=\delta(\,\sum^4_{i=1}\varepsilon_i-\epsilon)=
\delta(2T(u_--u_{3c}-\epsilon/2T)).
\label{bremdelta}
\eeq
The nucleon distribution functions in the term $\Xi[{\cal F}]$ in 
eq. (\ref{bremQ1}) have been rewritten in terms of the new 
direction cosines, the dimensionless momenta ($u_i$), and the initial-state 
nucleon degeneracy factors $\eta_{1,2}=\mu_{1,2}/T$:
\beq
{\cal F}_{1}=
\frac{e^{-(a_{1}\pr+ b\pr\gamma_1)}}{2\cosh(a_{1}\pr+ b\pr\gamma_1)}
\hspace{1cm}{\rm and}\hspace{1cm}
{\cal F}_{2}=
\frac{e^{-(a_{2}\pr- b\pr\gamma_1)}}{2\cosh(a_{2}\pr- b\pr\gamma_1)},
\eeq
where $a_{1,2}\pr =  a_{1,2}/2 = \frac{1}{2}(u_{+} + u_{-} - \eta_{1,2})$ 
and $b\pr = b/2 = (u_{+} u_{-})^{1/2}$.
Furthermore,
\beq
(1-{\cal F}_{3})=
\frac{e^{(c_{1}\pr+d\pr\gamma_c)}}{2\cosh(c_{1}\pr+d\pr\gamma_c)}
\hspace{1cm}{\rm and}\hspace{1cm}
(1-{\cal F}_{4})=
\frac{e^{(c_{2}\pr-d\pr\gamma_c)}}{2\cosh(c_{2}\pr-d\pr\gamma_c)},
\eeq
where $c_{1,2}\pr = c_{1,2}/2 = \frac{1}{2}(u_{+} + u_{3c} - \eta_{1,2})$ 
and $d\pr = d/2 = (u_{+} u_{3c})^{1/2}$.
${\cal F}_\nu$ and ${\cal F}_{\bar{\nu}}$, in contrast with the nucleon 
distribution functions, are independent
of angle; for a given set of thermodynamic conditions,
they remain functions of energy alone. While non-trivial, the 
integrations over $\gamma_1$ and $\gamma_c$ can be 
performed.  For example, the result for the $\gamma_1$ integration 
is of the form
\beq
\frac{1}{2\sqrt{B(B+1)}}
\ln\biggl[{(B - (1+2B)\xi^{2} + 2\xi{\sqrt B(B+1)
(\xi^{2} -1)}}\biggr], 
\eeq
where $B = \sinh^{2} a\pr$ and 
$\xi = \cosh b\pr \gamma_{1}$. With a proper evaluation of 
the integration limits and 
some algebra one can rewrite this result as
\beq
\frac{1}{2 \sinh a\pr \cosh a\pr} 
\ln\biggl[\frac{(1 + \cosh a \, \cosh b + \sinh a \, \sinh b)}{(
1 + \cosh a \, \cosh b - \sinh a \, \sinh b)}\biggr].
\eeq
Similar operations yield a result for the $\gamma_c$ integral 
in terms of $c$ and $d$.  In addition, eq. (\ref{bremdelta}) can be 
used  to eliminate the integral over $u_-$.  Collectively, these 
manipulations reveal that the differential $\nu_\mu$ bremsstrahlung 
emissivity at arbitrary neutron and proton degeneracy 
is simply a three-dimensional integral over $u_{+}$, $u_{3c}$, and 
$\varepsilon_{\bar{\nu}}$:
\beq
\frac{dQ}{d\varepsilon_\nu}=
Ks\zeta\,(1-{\cal F}_\nu)\,\varepsilon^3_\nu
\,\int d\varepsilon_{\bar{\nu}}\,du_+\,du_{3c}\,
(\varepsilon_{\bar{\nu}}/\epsilon)^2\,
u_+^{-1/2}e^{-\beta\epsilon/2}\,
\,\Phi(\epsilon,u_+,u_{3c})\,(1-{\cal F}_{\bar{\nu}}),
\label{totbrem2}
\eeq
where 
\beq
K=2G^2\left(\frac{m}{2\pi^2}\right)^{9/2}
\left(\frac{f}{m_\pi}\right)^4\,g_A^2\,T^{7/2},
\eeq
\beqa
\Phi(\epsilon,u_+,u_{3c})\,&=&\,\sinh^{-1}(f)\,{\rm ln}
\left[\left(\frac{1+\cosh(e_+)}{1+\cosh(e_-)}\right)
\left(\frac{\cosh(f)+\cosh(g_+)}
{\cosh(f)+\cosh(g_-)}\right)\right] \nonumber \\
&\times&\,\sinh^{-1}(j)\,{\rm ln}
\left[\left(\frac{1+\cosh(h_+)}{1+\cosh(h_-)}\right)
\left(\frac{\cosh(j)+\cosh(k_+)}{\cosh(j)+\cosh(k_-)}\right)
\right],
\label{phi}
\eeqa
and
\beqa
e_{\pm}&=&(u_+^{1/2}\pm u_-^{1/2})^2-\eta_2 \nonumber \\
f&=&u_++u_--\eta_1/2-\eta_2/2\nonumber \\
g_{\pm}&=& \pm 2(u_+u_-)^{1/2}-\eta_1/2+\eta_2/2\nonumber \\
h_{\pm}&=&(u_+^{1/2}\pm u_{3c}^{1/2})^2-\eta_2\nonumber \\
j&=&u_++u_{3c}-\eta_1/2-\eta_2/2\nonumber \\
k_{\pm}&=& \pm 2(u_+u_{3c})^{1/2}-\eta_1/2+\eta_2/2\,\,\,\,\,. 
\eeqa
Though $u_-$ has been integrated out via the energy-conserving 
delta function, it appears here in an attempt to make 
this expression more compact and should be read as 
$u_-=u_{3c}+\epsilon/2T$.  Importantly, if $\eta_1=\eta_2$ the 
right-hand term within both logarithmic terms in 
$\Phi(\epsilon,u_+,u_{3c})$ becomes unity. 

Using eq. \hspace{-8pt}(\ref{j_spec}), we can easily obtain the 
contribution to the Boltzmann equation due to nucleon-nucleon 
bremsstrahlung for arbitrary nucleon degeneracy, in asymmetric matter,
and including the full nucleon and neutrino Pauli blocking terms.
We find that
\beq
j_\nu
=K\pr s\zeta\,\int d\varepsilon_{\bar{\nu}}\,du_+\,du_{3c}\,
(\varepsilon_{\bar{\nu}}/\epsilon)^2\,
u_+^{-1/2}e^{-\beta\epsilon/2}\,
\,\Phi(\epsilon,u_+,u_{3c})\,(1-{\cal F}_{\bar{\nu}})
\eeq
where $K\pr=[(2\pi)^3/4\pi]K$.  The nucleon phase-space integrations 
above are identical in form for the $\nu_\mu\bar{\nu}_\mu$ absorption 
process, $\nu_\mu\bar{\nu}_\mu nn\rightarrow nn$. In this case, then, 
the primed energies are now associated with nucleons 1 and 2 in the above
manipulations and the incident nucleons (3 and 4) have unprimed energies. 
If we take the form derived above for the nucleon phase-space terms, the 
absorption channel ($\chi_\nu$) must then 
contain a factor of $e^{\beta\epsilon}$.  In addition, the blocking term,  
$(1-{\cal F}_{\bar{\nu}})$, becomes ${\cal F}_{\bar{\nu}}$. 
The Boltzmann equation for the evolution
of ${\cal F}_\nu$ in time is then, 
\beq
\frac{1}{c}\frac{\p {\cal F}_\nu}{\p t}
=K\pr s\zeta\int d\varepsilon_{\bar{\nu}}\,du_+\,du_{3c}\,
(\varepsilon_{\bar{\nu}}/\epsilon)^2\,
u_+^{-1/2}e^{-\beta\epsilon/2}\,
\,\Phi(\epsilon,u_+,u_{3c})\, 
\left\{(1-{\cal F}_\nu)(1-{\cal F}_{\bar{\nu}})-
{\cal F}_\nu{\cal F}_{\bar{\nu}}\,e^{\beta\epsilon}\right\}.
\label{workingbrem}
\eeq
For the neutron-neutron ($nn$) or proton-proton ($pp$) bremsstrahlung 
contribution, we simply set $s=1/4$ in eq. (\ref{workingbrem}) and use 
$\eta_1=\eta_2=\eta_n$ or $\eta_1=\eta_2=\eta_p$, respectively.
For the mixed nucleon ($np$) bremsstrahlung we set $s=1$, multiply 
eq. (\ref{workingbrem}) by $7/3$, and set $\eta_1=\eta_n$ and 
$\eta_2=\eta_p$.   While eqs. (\ref{totbrem2}) and (\ref{workingbrem}) 
may not appear symmetric in $\eta_1$ and $\eta_2$ the logarithmic terms 
conspire to ensure that the rates for both $np$ and $pn$ bremsstrahlung 
are identical, as they should be.  That is, it makes no difference whether 
we set $\eta_n$ or $\eta_p$ equal to $\eta_1$ or $\eta_2$.

Just as in \S\ref{subsec:pairpr}, in considering 
$e^+e^-\leftrightarrow\nu_\mu\bar{\nu}_\mu$, ${\cal F}_{\bar{\nu}}$
must be evolved simultaneously with ${\cal F}_\nu$.  In this case, however, 
the situation is simpler.  Suppose we wish to compare electron-positron 
annihilation with nucleon-nucleon bremsstrahlung by starting at $t=0$ with 
${\cal F}_{\bar{\nu}}={\cal F}_\nu=0$ over all energies.  We then solve 
eq. (\ref{workingpairpr}) and its ${\cal F}_{\bar{\nu}}$ counterpart at 
each timestep and at each energy. For $e^+e^-$ annihilation, ${\cal F}_\nu$ 
and ${\cal F}_{\bar{\nu}}$ will evolve differently; they will be visibly 
different at each timestep, because of the weighting of the
vector and axial-vector coupling constants which appear in the matrix 
element.  In contrast, eq. (\ref{workingbrem}) for bremsstrahlung must 
be solved only once. Since there is no difference in weighting between 
$\nu_\mu$ and $\bar{\nu}_\mu$, we can set 
${\cal F}_{\bar{\nu}}={\cal F}_\nu$ at every energy, at every 
timestep, as long as ${\cal F}_{\bar{\nu}}={\cal F}_\nu$
at $t=0$.  Of course, if we wish to consider 
${\cal F}_{\bar{\nu}}\neq{\cal F}_\nu$ initially, the two distributions 
would need to be evolved separately and simultaneously, coupled through 
the blocking and source terms on the right-hand side of the 
Boltzmann equation.

Equation (\ref{totbrem1}) can also be used to find the total volumetric 
$\nu_\mu\bar{\nu}_\mu$ pair emissivity.  To facilitate this we  replace 
$\varepsilon_\nu$ with $\epsilon$ and insert 
$\int\delta(\epsilon-(\varepsilon_\nu+\varepsilon_{\bar{\nu}}))d\epsilon$.
Assuming the neutrinos are radiated isotropically, we can use this delta 
function to do the integral over $d^3\vec{q}_{\bar{\nu}}$ and leave the 
total rate in terms of an integral over $\varepsilon_\nu$ from 
zero to $\epsilon$ and another over $\epsilon$ from zero to infinity. 
Momentarily ignoring neutrino blocking in the final state, the former 
can be integrated easily.  Making the same momentum, angle, and 
nucleon distribution function substitutions we used in deriving the 
single $\nu_\mu$ spectrum we can reduce the pair spectrum to an integral 
over $u_+$, $u_{3c}$, and $q=\epsilon/2T$.  We find that
\beq
Q_{\nu_\mu\bar{\nu}_\mu}=
Ds\zeta\,T^{8.5}\int dq\,du_{3c}\,du_+\,q^4e^{-q}\,u_+^{-1/2}
\,\Phi(\epsilon,u_+,u_{3c}),
\label{totspecpair}
\eeq
where
\beq
D=\frac{8}{15}\frac{G^2 g_A^2}{\sqrt{2}\,\pi^9}
\left(\frac{f}{m_\pi}\right)^4\,m^{9/2},
\eeq
and $\Phi(\epsilon,u_+,u_{3c})$ is defined in eq. (\ref{phi}). 
Note that eq. (\ref{totspecpair}) allows us to easily calculate the 
pair differential volumetric emissivity 
($dQ_{\nu_\mu\bar{\nu}_\mu}/d\epsilon$).  
For $Q^{nn}_{\nu_\mu\bar{\nu}_\mu}$ and 
$Q^{pp}_{\nu_\mu\bar{\nu}_\mu}$, $s=1/4$.  As with the single $\nu_\mu$ 
spectrum, for $Q^{np}_{\nu_\mu\bar{\nu}_\mu}$ multiply 
eq. (\ref{totspecpair}) by $7/3$ and set $s=1$.  Finally, 
$Q^{tot}_{\nu_\mu\bar{\nu}_\mu}$=$Q^{nn}_{\nu_\mu\bar{\nu}_\mu}$+
$Q^{pp}_{\nu_\mu\bar{\nu}_\mu}$+$Q^{np}_{\nu_\mu\bar{\nu}_\mu}$.

\subsubsection{The Non-Degenerate Nucleon Limit}

In the non-degenerate nucleon limit, the term 
${\cal F}_1{\cal F}_2(1-{\cal F}_3)(1-{\cal F}_4)$ reduces to
$e^{\eta_1}e^{\eta_2}e^{-2(u_++u_{-})}$ \cite{burrows_1999} which 
is independent of angle.  This tremendous simplification
allows for easy integration over $u_+$ and $u_{3c}$ in 
eqs. (\ref{totbrem2}), (\ref{workingbrem}), and 
{(\ref{totspecpair}). The total volumetric emissivity of a 
single $\nu_\mu\bar{\nu}_\mu$ pair in this limit,
ignoring $\nu_\mu$ and $\bar{\nu}_\mu$ blocking in the final
state, is \cite{burrows_1999}
\beq
Q_{\nu_\mu\bar{\nu}_\mu}\simeq
1.04\times10^{30}\zeta(X\rho_{14})^2
\left(\frac{T}{{\rm MeV}}\right)^{5.5}\,\,{\rm ergs\,cm^{-3}\,s^{-1}}.
\label{totalbremspec}
\eeq
For $nn$ and $pp$ bremsstrahlung, $X$ is the number fraction of 
neutrons ($X_n$) or protons ($X_p$), respectively.  For the mixed-nucleon 
process ($np$), $X^2$ becomes $(28/3)X_nX_p$. Figure \ref{fig:br_nn_tspec} 
compares the non-degenerate nucleon limit (eq. \ref{totalbremspec}) with 
the arbitrary nucleon degeneracy generalization (eq. \ref{totspecpair})
in the case of neutron-neutron ($nn$) bremsstrahlung, as a function of 
the neutron degeneracy $\eta_n=\mu_n/T$.  The filled square shows the 
degenerate limit obtained by ref. \cite{fsb_1975}.  Note that at 
$\eta_n\simeq0$, the fractional difference between the two is just 
$\sim12$\%.  At realistic neutron degeneracies
within the core ($\eta_n\sim2$), this difference approaches 30\%.

The single differential $\nu_\mu$ emissivity can be written in terms of 
the pair emissivity \cite{burrows_1999}:
\beqa
\frac{dQ}{d\varepsilon_\nu}&=&
C\left(\frac{Q_{\nu_\mu\bar{\nu}_\mu}}{T^4}\right)
\varepsilon_\nu^3\int_1^{\infty}\frac{e^{-2q_\nu x}}
{x^3}(x^2-x)^{1/2}dx \nonumber \\
&=&
C\left(\frac{Q_{\nu_\mu\bar{\nu}_\mu}}{T^4}\right)
\varepsilon_\nu^3\int_{q_\nu}^\infty\frac{e^{-q}}{q}
K_1(q)(q-q_\nu)^2\,dq\,\,,
\label{singlendbremspec}
\eeqa
where $C=2310/2048\simeq 1.128$, $q_\nu=\varepsilon_\nu/2T$, 
$q=\epsilon/2T$, and $K_1$ is the standard
modified Bessel function of imaginary argument.  
A useful fit to eq. (\ref{singlendbremspec}), good to better than 
3\% over the full range of 
relevant neutrino energies is \cite{burrows_1999}
\beq
\frac{dQ}{d\varepsilon_\nu}\sim
0.234\,\,\frac{Q_{\nu_\mu\bar{\nu}_\mu}}{T}
\,\left(\frac{\varepsilon_\nu}{T}\right)^{\hspace{-1.25pt}2.4}
e^{-1.1\varepsilon_\nu/T}.
\label{nubremspec}
\eeq
Using eq. (\ref{j_spec}), we obtain the 
contribution to the Boltzmann equation including Pauli blocking of 
$\nu_\mu$ and $\bar{\nu}_\mu$ neutrinos in the
final state: 
\beq
\frac{\p {\cal F}_\nu}{\p t}
={\cal C}s\zeta
\int_0^\infty\hspace{-5pt}d\varepsilon_{\bar{\nu}}
\,(\varepsilon^2_{\bar{\nu}}/\epsilon)
K_1\left(\frac{\beta\epsilon}{2}\right)e^{-\beta\epsilon/2}
\left\{(1-{\cal F}_\nu)(1-{\cal F}_{\bar{\nu}})-
{\cal F}_\nu{\cal F}_{\bar{\nu}}e^{\beta\epsilon}\right\}.
\label{boltzndbremspec}
\eeq
where
\beq
{\cal C}\,\,=\,\,\frac{G^2m^{4.5}}{\pi^{6.5}}
\left(\frac{f}{m_\pi}\right)^4\,g_A^2\, T^{2.5}\,e^{\eta_1}
\,e^{\eta_2}\,\,\simeq\,\, 
\frac{2G^2g_A^2}{\pi^{3.5}}\left(\frac{f}{m_\pi}\right)^4
\,\frac{m^{1.5}}{T^{.5}}\,n_1n_2\,.
\label{constantnd}
\eeq
In obtaining eq. (\ref{constantnd}), we have used the 
thermodynamic identity in the non-degenerate limit,
\beq
e^{\eta_i}=\left(\frac{2\pi}{mT}\right)^{3/2}\frac{n_i}{2},
\eeq
where $n$ is the number density of nucleons considered and $i$ 
is 1 or 2 for neutrons or protons, depending on which nucleon 
bremsstrahlung process is considered.

\section{Results}
\label{sec:results}

The numerical algorithm we have developed accepts arbitrary 
initial $\nu_\mu$ and $\bar{\nu}_\mu$ phase-space distributions.
Using the scattering formalism developed in the previous section, 
we evolve two initial distribution functions: (1) a broad Gaussian 
in energy centered at 40 MeV with a maximum of ${\cal F}_\nu=0.80$ 
and a full-width at half-maximum of $\sim$28.6 MeV, and
(2) a Fermi-Dirac distribution at a temperature 2$\,\times$ 
the temperature of the surrounding matter and
with zero chemical potential. While the former is unphysical
in the context of supernova calculations, it illustrates the 
effects of blocking on both the average energy 
transfer and the rates for each scattering process.  
Furthermore, its evolution is more dynamic than the Fermi-Dirac 
distribution. As a result, the way in which the distribution is 
spread and shifted in time is more apparent.  The essential 
differences between the two processes are then more easily gleaned.
The latter initial distribution is motivated by consideration of 
the environment within the core of a supernova.
The $\nu_\mu$ and $\bar{\nu}_\mu$ distribution functions, having 
been generated as pairs via $e^+e^-\leftrightarrow\nu_\mu\bar{\nu}_\mu$ 
and nucleon-nucleon bremsstrahlung should have approximately 
zero chemical potential.  Furthermore, even in the dense core, the 
$\nu_\mu$s will diffuse outward in radius and,
hence, from higher to lower temperatures.  By starting with a 
Fermi-Dirac distribution at twice the temperature of
the matter at that radius, we learn more about how equilibration 
might effect the emergent $\nu_\mu$ spectrum in an actual collapse 
or protoneutron star cooling calculation.

For the production and emission processes, we start with zero 
neutrino occupancy and let each build to an equilibrium distribution 
of $\nu_\mu$s and $\bar{\nu}_\mu$s.  As a check to the calculation,
the asymptotic distribution should be Fermi-Dirac at the temperature 
of the ambient matter with zero neutrino chemical potential. 
Throughout these simulations, we take the factor $\zeta$ in eq.
(\ref{workingbrem}) for nucleon-nucleon bremsstrahlung to be 0.5. 
(This factor represents our ignorance of the importance 
of the nucleon momentum transfer terms.)

We repeat these calculations for four temperature, density, and 
composition points (StarA, StarB, StarC, and StarD) taken from the 
one-dimensional collapse calculation profile, {\it Star} 
\cite{bhf_1995},  corresponding to four radii 
below the shock ($\sim\hspace{-1pt}80$ km).  Roughly, these points 
have densities $10^{14}$, $10^{13}$, $10^{12}$, 
and $10^{11}$ g cm$^{-3}$.  The actual numbers are shown in 
Table \ref{tab:star}.

\subsection{Scattering}

Figures \ref{fig:ns_star36_fnu} and \ref{fig:es_star36_fnu} 
show the evolution of a Gaussian distribution
at $t=0$ to an equilibrium Fermi-Dirac distribution at the 
temperature of the surrounding matter due to $\nu_\mu$-neutron 
($\nu_\mu n$) and $\nu_\mu$-electron ($\nu_\mu e^-$) scattering, 
respectively. The equilibrium distribution has a non-zero neutrino 
chemical potential set by the initial total number of 
$\nu_\mu$s, which is conserved to better than .001\% throughout 
the calculation.  Multiple curves on each plot show snapshots of 
${\cal F}_\nu$ in time from $t=0$ to 1000 microseconds ($\mu$s).
Both calculations were carried out at the thermodynamic point StarB 
whose characteristics are shown in Table \ref{tab:star}. StarB is 
indicative of the core of a supernova, a region of moderate to high 
temperatures ($T\sim15$ MeV) and densities of $\sim10^{13}$ g cm$^{-3}$.
These two figures illustrate the fundamental differences between 
$\nu_\mu e^-$ and $\nu_\mu n$ scattering as thermalization processes.  
Curve A in Fig. \ref{fig:ns_star36_fnu} and  curve C
in Fig. \ref{fig:es_star36_fnu} indicate that
at high $\nu_\mu$ energies ($\varepsilon_\nu\gtrsim$ 30 MeV) 
$\nu_\mu n$ scattering is a much more effective thermalization 
mechanism. At $\varepsilon_\nu\simeq40$ MeV both curves show the 
distribution is within $\sim$30\% of equilibrium.  Importantly, 
however, curve A is at 0.33 $\mu$s for $\nu_\mu n$ scattering 
whereas curve C is at 3.30 $\mu$s for $\nu_\mu e^-$ scattering.  
Curve C, in Fig. \ref{fig:ns_star36_fnu} for $\nu_\mu n$ scattering, 
also at $t=3.30$ $\mu$s, shows that above $\sim$25 MeV 
the distribution has almost equilibrated.  For $\nu_\mu e^-$ 
scattering, similar evolution at high neutrino energies takes 
approximately 25 $\mu$s. These simple estimates reveal
that $\nu_\mu n$ scattering is about 10 times faster than 
$\nu_\mu e^-$ scattering at equilibrating $\nu_\mu$s with 
energies greater than approximately 25 MeV.  

This situation is reversed at low $\varepsilon_\nu$s. Comparing 
curve E at $t=33.0$ $\mu$s in both Fig. \ref{fig:ns_star36_fnu} 
and Fig. \ref{fig:es_star36_fnu}, we can see that at $\sim10$ MeV 
both distributions have filled to approximately the same percentage 
of the asymptotic, equilibrium ${\cal F}_\nu$.  However, below 
$\varepsilon_\nu\sim$8 MeV, $\nu_\mu n$ scattering has not filled 
${\cal F}_\nu$ to the extent $\nu_\mu e^-$ scattering has.
In fact, the rate at which these low energy states are filled by 
$\nu_\mu n$ scattering is very low; the energy transfer ($\omega$) 
is much smaller than the incident $\nu_\mu$ energy.  In this regime, 
the Fokker-Planck approximation for the time evolution of ${\cal F}_\nu$ 
in energy space may be applicable.  In marked contrast, 
Fig. \ref{fig:es_star36_fnu} indicates how effective $\nu_\mu e^-$ 
scattering is at filling the lowest $\varepsilon_\nu$ states.  
Curves F from Figs. \ref{fig:ns_star36_fnu} and
\ref{fig:es_star36_fnu}, taken at 1000 $\mu$s, 
show that though the distribution has reached equilibrium via 
$\nu_\mu e^-$ scattering, for $\nu_\mu n$ scattering the very 
lowest energy states remain unfilled. For each of the four points 
in the {\it Star} profile we consider, $\nu_\mu n$ scattering
dominates at high energies ($\gtrsim 20$ MeV), whereas $\nu_\mu e^-$ 
scattering dominates at low $\nu_\mu$ energies
($\lesssim10$ MeV) and particularly for $\varepsilon_\nu\lesssim3$ MeV.

Figures \ref{fig:ns_star68_newfd200_fnu} and 
\ref{fig:es_star68_newfd200_fnu} depict the evolution of ${\cal F}_\nu$ 
via $\nu_\mu n$ and $\nu_\mu e^-$ scattering, respectively, for an 
initial Fermi-Dirac distribution at 2$\times$ the temperature of the 
surrounding neutrons and electrons and with zero neutrino chemical 
potential. This calculation was carried out at 
StarC (see Table \ref{tab:star}), which is representative of 
the outer core, in the semi-transparent regime, where the neutrinos 
begin to decouple from the matter (near the neutrinosphere).
The same systematics highlighted in the discussion of the evolution 
of the initial Gaussian distribution for StarB are borne out in these 
figures.  Curves A and B on both plots, denoting 0.10 and 0.33 milliseconds 
(ms) of elapsed time, respectively, confirm that above 
$\varepsilon_\nu\sim15$ MeV $\nu_\mu n$ scattering dominates thermalization.

Figures \ref{fig:ns_star68_newfd200_winout} and 
\ref{fig:es_star68_newfd200_winout} show $\langle\omega\rangle_{in}$ and
$\langle\omega\rangle_{out}$, as defined in eqs. (\ref{win}) and (\ref{wout}), 
for $\nu_\mu n$ scattering and $\nu_\mu e^-$ scattering, respectively. 
The separate curves portray the evolution in time of the thermal average
energy transfers as the distributions evolve to equilibrium 
(cf. Figs. \ref{fig:ns_star68_newfd200_fnu} and 
\ref{fig:es_star68_newfd200_fnu}).  As one would expect 
from kinematic arguments, the magnitudes of both $\langle\omega\rangle_{in}$ 
and $\langle\omega\rangle_{out}$ for $\nu_\mu n$ scattering 
are much less than those for $\nu_\mu e^-$ scattering.  Though the energy
transfers are much smaller, even at the highest energies, $\nu_\mu n$ 
scattering still dominates $\nu_\mu e^-$ scattering
in thermalizing the $\nu_\mu$ distribution because the rate for scattering 
is so much larger.  At low neutrino energies, however, both average energy 
transfers for neutron scattering go to zero, whereas they 
approach large negative values ($\sim-20$ MeV) for electron scattering.  At 
these low energies, the fact that the rate for $\nu_\mu n$ scattering is 
larger than for $\nu_\mu e^-$ scattering fails to compensate for the vanishing 
energy transfer.  For example, at $\varepsilon_\nu=3$ MeV and $t=33$ ms, 
the energy transfer for $\nu_\mu e^-$ scattering
is more than 100 times that for $\nu_\mu n$ scattering.  

In order to fold in information about both the rate of scattering and the 
average thermal energy transfer, we plot $\Gamma_D$ and $\Gamma_E$ 
(eqs. \ref{gammad} and \ref{gammae}) in 
Fig. \ref{fig:nses_allstar_newfd400_tub} for all four points considered 
in the {\it Star} profile.  We show here a snapshot of the rates for both 
scattering processes for a Fermi-Dirac distribution initially at twice the 
local matter temperature, with zero neutrino chemical potential.
Note that the spikes in $\Gamma_D$ indicate the neutrino energy at which 
$\langle\omega\rangle_{out}=0$ (cf. Figs. \ref{fig:ns_star68_newfd200_winout} 
and \ref{fig:es_star68_newfd200_winout}). In general, we find that as 
$\varepsilon_\nu\rightarrow0$,  $\Gamma_D$ and $\Gamma_E$ go to zero for 
$\nu_\mu$-neutron scattering, whereas $\Gamma_D$ approaches a constant 
and $\Gamma_E$ gets very large  for $\nu_\mu e^-$ scattering 
\cite{tubbs_scatter}.  This is a consequence of the fact that, regardless 
of ${\cal F}_\nu$, $\langle\omega\rangle_{out}\rightarrow0$ for 
$\nu_\mu n$ scattering as $\varepsilon_\nu\rightarrow0$, as shown in 
Fig. \ref{fig:ns_star68_newfd200_winout}.  For $\nu_\mu e^-$ scattering 
the situation is different.  As Fig. \ref{fig:es_star68_newfd200_winout} 
reveals,  $\langle\omega\rangle_{out}$ approaches $\sim-20$ MeV at 
$\varepsilon_\nu=0$. As expected from our analysis of the 
evolution of ${\cal F}_\nu$ in Figs. \ref{fig:ns_star68_newfd200_fnu} 
and \ref{fig:es_star68_newfd200_fnu}, at approximately 40 MeV the 
thermalization rate for $\nu_\mu n$ scattering for StarB is about an 
order of magnitude greater than that for $\nu_\mu e^-$ scattering. 
Specifically, the $\Gamma_D$'s cross at $\sim$15 MeV, whereas the 
$\Gamma_E$'s cross at $\sim$20 MeV.  Below these energies, 
both $\nu_\mu n$ rates drop off precipitously as a consequence of the 
fact that $\langle\omega\rangle_{out}\rightarrow0$.  Below 
$\varepsilon_\nu\sim5$ MeV, the thermalization rate for $\nu_\mu e^-$ 
scattering dominates by 2-5 orders of magnitude. As evidenced by the 
other panels in Fig. \ref{fig:nses_allstar_newfd400_tub},
this same trend holds in the other regions of the stellar profile. 
In general, the rates drop over the whole energy range for both processes 
as the density and temperature decrease, but the same systematics
hold.  In fact, for StarA, StarC, and StarD the $\Gamma_E$ and $\Gamma_D$ 
crossing points for both processes are lower
than those for StarB.  As a result of the higher temperature at this 
radius ($T\simeq14.5$ MeV) $\nu_\mu e^-$ scattering
is important in thermalizing slightly higher energy neutrinos than 
at the other radii.  For StarC and StarD, specifically,
both rates cross at neutrino energies less than 12 MeV.

These results demonstrate that $\nu_\mu$-nucleon scattering is an 
important thermalization process from the dense core 
through the semi-transparent regime for $\nu_\mu$s with energies 
greater than approximately 15 MeV.  The addition of this energy transfer 
mechanism implies that the $\nu_\mu$s stay energetically coupled to the 
surrounding matter longer than has been previously estimated 
\cite{burrows_mazurek}.  We can approximate the radius at which
the $\nu_\mu$s energetically decouple from the matter 
(the $E_\mu$-sphere) \cite{burrows_mazurek} by observing when the 
diffusion timescale is approximately equal to the
equilibration timescale given by $\Gamma_D^{-1}=\tau_D$, as defined 
in eq. (\ref{gammad}).  Using this crude approximation we find that 
by including $\nu_\mu$-nucleon energy transfer
the $E_\mu$-sphere is pushed outward in radius by 
approximately 3 kilometers.  This difference in radius corresponds 
to a 1-2 MeV drop in the matter temperature in the model {\it Star}.  
The average energy of the emergent spectrum is roughly correlated with
the local matter temperature of the $E_\mu$-sphere.  
Therefore, we conclude that $\nu_\mu$-nucleon energy transfer in full 
transport calculations will likely soften the emergent $\nu_\mu$ spectrum.

\subsection{Emission and Absorption}

Figure \ref{fig:radcomp_star} shows the total integrated volumetric 
emissivity as a function of radius in the model {\it Star} for 
nucleon-nucleon bremsstrahlung in the non-degenerate nucleon limit 
(eq. \ref{totalbremspec}), its generalization for  arbitrary nucleon 
degeneracy (eq. \ref{totspecpair}), and the emissivity for $e^+e^-$ 
annihilation (eq. \ref{eetspec}).  Note that not one of these expressions 
contains neutrino blocking terms and that the general bremsstrahlung rate 
crosses that for $e^+e^-$ annihilation at $\sim 23$ kilometers where 
$\rho\simeq6\times10^{12}$ g cm$^{-3}$, $T\simeq11$ MeV, and $Y_e\simeq0.13$.  
While this plot gives a general idea of where $e^+e^-$ annihilation should 
begin to compete with nucleon-nucleon bremsstrahlung, it
fails to include the differential nature of the production in energy.  
In addition, it does not include absorption or blocking effects, which 
quantitatively alter the relative strength of the emission.

To begin to understand the import of these terms and the character of 
each pair production process, we include Figs. \ref{fig:br_star68_fnu} 
and \ref{fig:pp_star68_fnu}, which show the time evolution of 
${\cal F}_\nu$ via nucleon-nucleon bremsstrahlung and electron-positron 
annihilation, respectively, for the point StarC, initialized with zero
$\nu_\mu$ and $\bar{\nu}_\mu$ phase-space occupancies.  The final equilibrium 
distribution is Fermi-Dirac at the temperature of the surrounding matter, 
with zero neutrino chemical potential.  Comparing curve C on both graphs,
which marks 10.0 milliseconds (ms) of elapsed time, one can see that 
bremsstrahlung dominates production below $\sim15$ MeV. Indeed, 
bremsstrahlung overshoots its equilibrium distribution at energies below 
10 MeV before finally filling the higher $\varepsilon_\nu$ states.  
In contrast, electron-positron annihilation fills the higher states 
first and moves slowly toward the low-lying neutrino energies, taking 
a factor of 10 more time  at this thermodynamic point to reach equilibrium.  

In Figs. \ref{fig:brpp_star36_tin} and \ref{fig:brpp_star36_tout}, we 
plot $\Gamma_{in}$ and $\Gamma_{out}$, as defined in 
eqs. (\ref{gammain}) and (\ref{gammaout}), for both production 
processes at the point StarB.  As one would predict from our 
simple observations of the time evolution of ${\cal F}_\nu$, 
the bremsstrahlung rates are much faster ($\sim$2 orders of magnitude) 
than the $e^+e^-$ annihilation rates at low neutrino energies.  
At StarC, $e^+e^-$ annihilation competes with bremsstrahlung above 
$\varepsilon_\nu\sim15$ MeV.  For StarB, however, at a matter 
density an order of magnitude greater than that for StarC, 
the energy at which nucleon-nucleon bremsstrahlung becomes more 
important than $e^+e^-$ annihilation is $\sim60$ MeV.  In this 
regime, where $T\sim12-14$ MeV and $\rho\sim10^{13}$ g cm$^{-3}$, 
we find that bremsstrahlung dominates neutrino pair-production via
electron-positron annihilation.  A close look at the evolution of 
the total thermal average neutrino energy 
($\langle\varepsilon_\nu\rangle$) reveals that ${\cal F}_\nu$ 
reaches its asymptotic equilibrium distribution
via nucleon-nucleon bremsstrahlung in $\sim1$ ms.  Electron-positron 
annihilation takes $\sim50$ ms to fill all but the very lowest energy 
states. This trend continues as the matter becomes more dense.  
For StarA, well beneath the neutrinospheres at 
$\rho\sim10^{14}$ g cm$^{-3}$, the rates for bremsstrahlung and 
electron-positron annihilation never cross.  In fact, the former 
produces an equilibrium Fermi sea of $\nu_\mu$'s in $\sim 50$ $\mu$s, 
whereas the latter takes $\sim10^3$ seconds.  This difference of 8 
orders of magnitude in timescale, however, is a bit misleading. 
Similar to $\nu_\mu n$ scattering, $e^+e^-$ annihilation has trouble 
filling only the very lowest neutrino energy states.  
In actuality, at the highest energies, both $\Gamma_{in}$ and 
$\Gamma_{out}$ for $e^+e^-$ annihilation come within 3-4 
orders of magnitude of the rates for bremsstrahlung at the same energy.  
Still, the difference is striking.  As the temperature drops from 
StarB (14 MeV) to StarA (10 MeV) and the density increases by an 
order of magnitude, $\eta_e$ goes from 3.79 to 15.75.  Consequently, 
Pauli blocking of electrons in the final state suppresses
the process $\nu_\mu\bar{\nu}_\mu\rightarrow e^+e^-$, and the phase-space 
density of positrons is depleted to such an
extent that $e^+e^-\rightarrow\nu_\mu\bar{\nu}_\mu$ is suppressed as well.  
We conclude that beneath the neutrinospheres
and specifically for $\rho\sim10^{13}$ g cm$^{-3}$, nucleon-nucleon 
bremsstrahlung is the primary and dominant $\nu_\mu\bar{\nu}_\mu$ source.   
Near the neutrinosphere, within the gain region and behind the shock, 
between 30 km and 60 km at  $\rho\sim10^{12}$ g cm$^{-3}$ and 
$T\sim6-8$ MeV, bremsstrahlung competes with $e^+e^-$ annihilation 
at all neutrino energies and is the primary production process for 
the low-lying $\varepsilon_\nu$ and $\varepsilon_{\bar{\nu}}$ states.

The addition of nucleon-nucleon bremsstrahlung will have quantitative 
implications for the $\nu_\mu$ and $\nu_\tau$ emergent spectra.  
Specifically, they should be softer and brighter. 
Burrows et al. \cite{burrows_1999} confirm this with their study of 
static supernova and protoneutron star atmospheres, having included 
nucleon-nucleon bremsstrahlung in the non-degenerate limit.  In addition 
to observing a systematic softening, they also find that 
the $\nu_\mu$ spectrum is a factor of 2 more luminous 
at $\varepsilon_\nu=10$ MeV.

\section{Summary and Conclusions}
\label{sec:conclusions}

Our results for equilibration via $\nu_\mu$-electron scattering and
$\nu_\mu$-nucleon scattering indicate that the latter competes with 
or dominates the former as a thermalizer for neutrino energies 
$\gtrsim10$ MeV for $\rho\gtrsim1\times10^{11}$ g cm$^{-3}$ at all 
temperatures. At neutrino energies $\gtrsim30$ MeV the difference at 
all densities and temperatures is approximately an order
of magnitude. For the production and absorption processes, we find 
that nucleon-nucleon bremsstrahlung, at the average energy of an 
equilibrium Fermi-Dirac distribution at the local temperature, 
is 5 and 2 orders of magnitude faster than $e^+e^-$ annihilation 
at StarA ($T\sim10$ MeV, $\rho\sim10^{14}$ g cm$^{-3}$) and StarB 
($T\sim15$ MeV, $\rho\sim10^{13}$ g cm$^{-3}$), respectively.  Only 
for $\rho\sim10^{12}$ g cm$^{-3}$ and $T\sim6$ MeV does 
$e^+e^-\leftrightarrow\nu_\mu\bar{\nu}_\mu$ begin to compete with 
bremsstrahlung at all energies.  We conclude from this study that 
the emergent $\nu_\mu$ and $\nu_\tau$ spectrum is (1) brighter and 
(2) softer than previously estimated.  The former results from the 
inclusion of the new pair emission process, nucleon-nucleon bremsstrahlung.
The latter is a consequence of both the increased energy coupling 
between the nuclear and neutrino fluids through $\nu_\mu$-nucleon 
scattering and the fact that bremsstrahlung dominates $e^+e^-$ 
annihilation near the neutrinospheres at the lowest neutrino energies.
While the full transport problem, including $\nu_\mu$-nucleon scattering 
energy redistribution and nucleon-nucleon bremsstrahlung, must be solved 
in order to delineate precisely what consequences these 
processes have for the emergent $\nu_\mu$ spectrum, these calculations 
demonstrate that they should not be omitted.

\section{Acknowledgments}

The authors thank Sanjay Reddy for helpful correspondence.  
A.B. and T.A.T. acknowledge support under NSF Grant No. AST96-14794 and 
J.E.H. acknowledges funding from the 
Fundac\~{a}o de Amparo a Pesquisa do Estado de S\~{a}o Paulo.

\appendix
\section{Neutrino-electron scattering}
\label{app:escatt}

Each of the retarded polarization functions in eqs. (\ref{s1}-\ref{s3}) 
can be written in terms of one-dimensional integrals over 
electron energy ($\varepsilon_e$),  which we label $I_n$ \cite{reddy_1998};  
\beq
{\rm Im}\Pi_L^R(q,\omega)\,=
\,\frac{q_\mu^2}{2\pi|q|^3}\left[I_2+\omega I_1+
\frac{q_\mu^2}{4}I_0\right],
\eeq
\beq
{\rm Im}\Pi_T^R(q,\omega)\,=
\,\frac{q_\mu^2}{4\pi|q|^3}\left[I_2+\omega I_1+
\left(\frac{q_\mu^2}{4}+\frac{q^2}{2}+
m^2\frac{q^2}{q_\mu^2}\right)I_0\right],
\eeq
\beq
{\rm Im}\Pi_A^R(q,\omega)\,=\,\frac{m^2}{2\pi|q|}I_0,
\eeq
and
\beq
{\rm Im}\Pi_{VA}^R(q,\omega)\,=
\,\frac{q_\mu^2}{8\pi|q|^3}\left[\omega I_0+2I_1\right].
\eeq
The authors of \cite{reddy_1998} were able to express the 
$I_n$'s in terms of polylogarithmic integrals such that
\beq
I_0\,=\,Tz\left(1-\frac{\xi_1}{z}\right),
\label{I0}
\eeq
\beq
I_1\,=\,T^2z\left(\eta_e-\frac{z}{2}-\frac{\xi_2}{z}-
\frac{e_-\xi_1}{zT}\right), 
\label{I1}
\eeq
and
\beq
I_2\,=\,T^3z\left(\eta_e^2-z\eta_e+\frac{\pi^2}{3}+
\frac{z^2}{3}+2\frac{\xi_3}{z}-
2\frac{e_-\xi_2}{Tz}+\frac{e_-^2\xi_1}{T^2z}\right),
\label{I2}
\eeq
where $\eta_e=\mu_e/T$ is the electron degeneracy, 
$z=\beta\omega$, $\omega$ is the energy transfer, and 
\beq
e_-=-\frac{\omega}{2}+\frac{q}{2}\sqrt{1-4\frac{m^2}{q_\mu^2}}.
\eeq
In eqs. (\ref{I0}-\ref{I2}), the $\xi_n$'s are differences 
between polylogarithmic integrals;
$\xi_n={\rm Li}_n(-\alpha_1)-{\rm Li}(-\alpha_2)$, where
\beq
{\rm Li}_n(y)=\int_0^y\frac{{\rm Li}_{n-1}(x)}{x}\,dx, 
\eeq
and Li$_1(x)={\rm ln}(1-x)$.  The arguments necessary for 
computing the integrals are 
$\alpha_1={\rm exp}[\beta(e_-+\omega)-\eta_e]$ and 
$\alpha_2={\rm exp}(\beta e_--\eta_e)$.

\section{electron-positron annihilation}
\label{app:pairpr}

The production kernel is defined by 
\beq
R^p(\varepsilon_\nu,\varepsilon_{\bar{\nu}},\cos\theta)=
\frac{1}{2\varepsilon_\nu\varepsilon_{\bar{\nu}}}
\int\frac{d^3\vec{p}}{(2\pi)^3 2\varepsilon}
\frac{d^3\vec{p}^{\,\prime}}{(2\pi)^3 2\varepsilon\pr}
{\cal F}_{e^-}{\cal F}_{e^+}\left(\frac{1}{4}\sum_s|{\cal M}|^2\right)
\,(2\pi)^4\delta^4({\bf {\rm{P}}}).
\eeq
The differential production spectrum for final state $\nu_\mu$s can 
then be written as \cite{bruenn_1985}
\beq
\frac{dQ}{d\varepsilon_\nu}=(1-{\cal F}_\nu)
\frac{\varepsilon_\nu^3}{(2\pi)^6}\,\int\,d\Omega\int_0^\infty
\varepsilon_{\bar{\nu}}^2\,d\varepsilon_{\bar{\nu}}
\,\int_{-1}^1\,d\mu\pr\,\int_0^{2\pi}\,d\phi\,
R^p(\varepsilon_\nu,\varepsilon_{\bar{\nu}},\cos\theta)
\,(1-{\cal F}_{\bar{\nu}})\,\,,
\label{spectrumR}
\eeq
where $d\Omega$ is the differential solid angle for the $\nu_\mu$ 
neutrino, $\mu\pr=\cos\theta\pr$ is the cosine of the $\bar{\nu}_\mu$ 
angular coordinate, and $\phi$ is the azimuthal angle between 
$\nu_\mu$ and $\bar{\nu}_\mu$. Expanding the production kernel in a 
Legendre series in the scattering angle, 
$\cos\theta=\mu\mu\pr+[(1-\mu^2)(1-\mu^{\prime\,2})]^{1/2}\cos\phi$,
\beq
R^p(\varepsilon_\nu,\varepsilon_{\bar{\nu}},\cos\theta)=
\frac{1}{2}\sum_l(2l+1)
\Phi^p_l(\varepsilon_\nu,\varepsilon_{\bar{\nu}})P_l(\cos\theta)
\sim\frac{1}{2}\Phi^p_0(\varepsilon_\nu,\varepsilon_{\bar{\nu}})+\frac{3}{2}
\Phi^p_1(\varepsilon_\nu,\varepsilon_{\bar{\nu}})\cos\theta.
\label{expansion}
\eeq
$\Phi^p_0$, in eqs. (\ref{spectrum}) and (\ref{expansion}), is 
given by \cite{bruenn_1985,yueh_76}
\beq
\Phi_0^p(\varepsilon_\nu,\varepsilon_{\bar{\nu}})=\frac{G^2}{\pi}
\int_0^{\varepsilon_\nu+\varepsilon_{\bar{\nu}}}
\hspace{-12pt}d\varepsilon\,\,{\cal {F}}_{e^-}
{\cal {F}}_{e^+}\,
H_0(\varepsilon_\nu,\varepsilon_{\bar{\nu},}\varepsilon),
\label{phi0}
\eeq
where ${\cal {F}}_{e^+}$ is a function of 
$\varepsilon\pr(=\varepsilon_\nu+\varepsilon_{\bar{\nu}}-\varepsilon)$ and 
\beq
H_0(\varepsilon_\nu,\varepsilon_{\bar{\nu}},\varepsilon)=(V+A)^2\,
J_0^I(\varepsilon_\nu,\varepsilon_{\bar{\nu}},\varepsilon)+(V-A)^2\,
J_0^{II}(\varepsilon_\nu,\varepsilon_{\bar{\nu}},\varepsilon)\,\,.
\label{polynomials}
\eeq
Each $J_0$ in eq. (\ref{polynomials}) is a polynomial in 
$\varepsilon_\nu$, $\varepsilon_{\bar{\nu}}$, 
and $\varepsilon$ of dimension [energy]. They are related to each 
other by \cite{bruenn_1985}
\beq
J_0^I(\varepsilon_\nu,\varepsilon_{\bar{\nu}},\varepsilon)=
J_0^{II}(\varepsilon_{\bar{\nu}},\varepsilon_\nu,\varepsilon)\,\,.
\label{interchangejs}
\eeq
Both $J_0^I$ and $J_0^{II}$ can be found in ref. \cite{bruenn_1985}
(correcting for the typo in their eq. C67).
From eqs. (\ref{spectrum}) and (\ref{polynomials}) we see that the 
differences between the spectra for $\nu_\mu$s and $\bar{\nu}_\mu$s 
for a given temperature and electron degeneracy ($\eta_e$) arise solely 
from the relative weighting constants $(V+A)^2$ and $(V-A)^2$ in 
eq. (\ref{polynomials}) for $J_0^I$ and $J_0^{II}$, respectively.
Indeed, in this approximation the same can be said for the difference 
in the spectrum between $\nu_e$ and $\nu_\mu$ neutrinos.  

\pagebreak

\vspace{2cm}
\begin{table}
\caption{Radius ($R$), temperature ($T$), density ($\rho$), lepton
fraction ($Y_e$), and degeneracy factors ($\eta=\mu/T$)
for neutrons, protons, and electrons for four points from the
model, {\it Star}, a one-dimensional core-collapse calculation
evolved through collapse for 0.24 seconds. At this point in the
core evolution, the shock is at about 80 kilometers Burrows, Hayes, and Fryxell (1995).\\
\vspace*{.1cm}}
\begin{tabular}{cccccccc}
\hspace{1cm}Label & $R$ (km) &$\rho$ (g cm$^{-3}$) & $T$(MeV) & 
$Y_e$ & $\eta_n$ & $\eta_p$ & $\eta_e$ \hspace{1cm} \\
\tableline
\vspace{-.2cm} \\
\hspace{1cm}StarA & 10.75 & $1.281\times10^{14}$ 
& 10.56 & 0.2752 & 2.37 & 0.70  & 15.75 \hspace{1cm} \\
\vspace{-.2cm} \\
\hspace{1cm}StarB & 18.75 & $1.023\times10^{13}$ 
& 14.51 & 0.2021 & -1.62 & -3.04 & 3.79 \hspace{1cm} \\
\vspace{-.2cm} \\
\hspace{1cm}StarC & 34.75 & $1.082\times10^{12}$ 
& 6.139 & 0.0907 & -2.48 & -4.81 & 3.03 \hspace{1cm} \\
\vspace{-.2cm} \\
\hspace{1cm}StarD & 49.75 & $1.071\times10^{11}$ 
& 4.527 & 0.1671 & -4.45 & -6.06 & 1.93 \hspace{1cm} \\
\end{tabular}
\label{tab:star}
\end{table}

\pagebreak

\begin{figure} 
\vspace*{7.50in}
\hbox to\hsize{\hfill\includegraphics{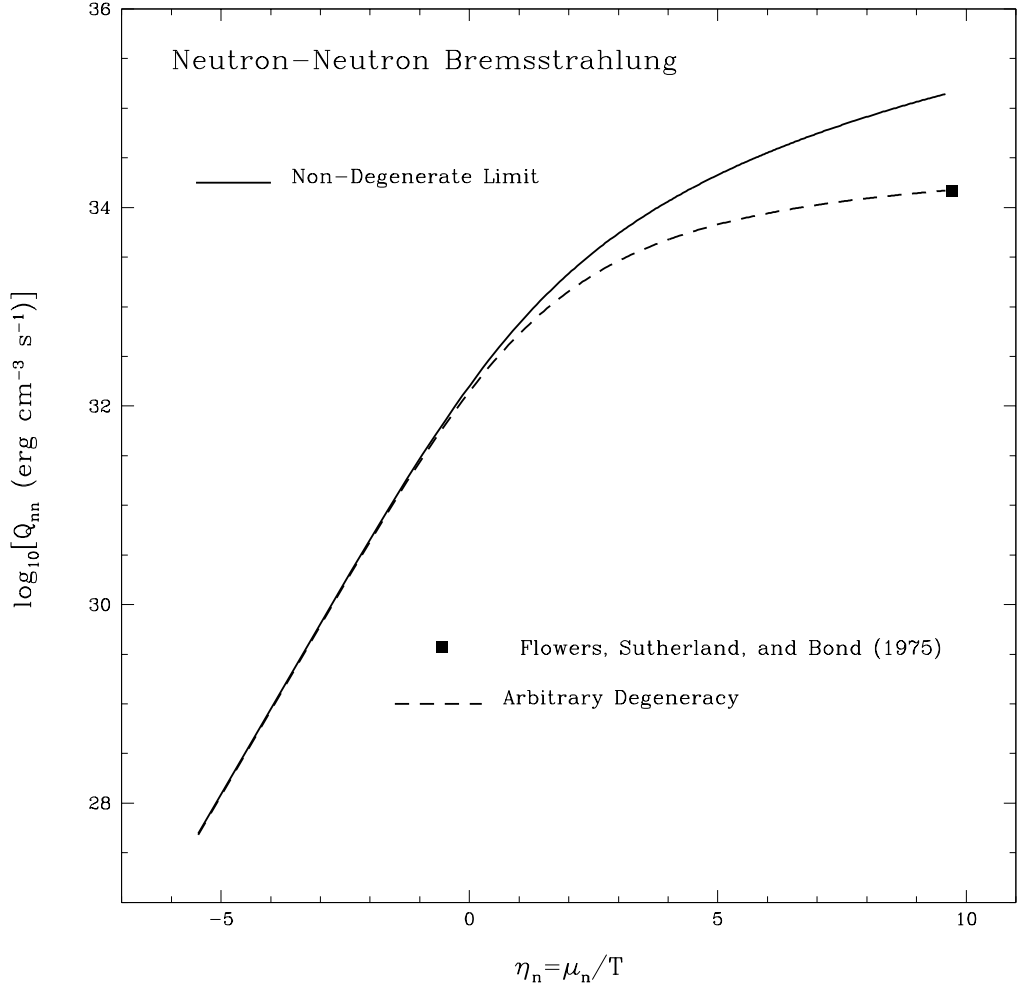}\kern+6in\hfill}
\caption{
The total volumetric emissivity due to neutron-neutron 
bremsstrahlung ($Q_{nn}$) in ergs cm$^{-3}$ s$^{-1}$ in
the non-degenerate neutron limit (solid line, eq. \ref{totalbremspec}) 
and at arbitrary nucleon degeneracy (dashed line, eq. \ref{totspecpair}) 
for $T=6$ MeV, $Y_e=0.0$, and for a range of densities from 
$5\times10^{10}$ g cm$^{-3}$ to nuclear density 
($\sim2.68\times10^{14}$ g cm$^{-3}$).  The filled box denotes
the degenerate neutron limit obtained by Flowers, Sutherland, and Bond (1975).}
\label{fig:br_nn_tspec}
\end{figure}

\begin{figure} 
\vspace*{7.50in}
\hbox to\hsize{\hfill\includegraphics{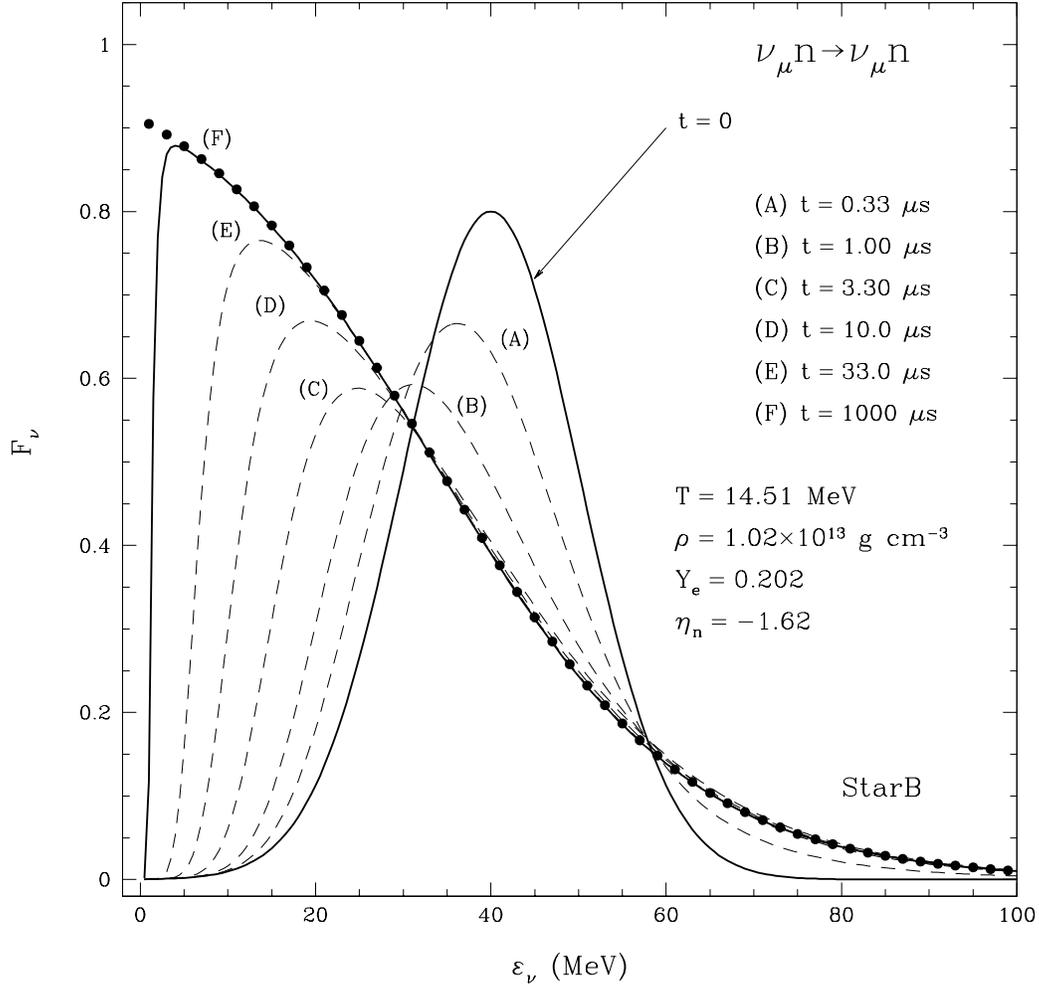}\kern+6in\hfill}
\caption{
The time evolution via $\nu_\mu$-neutron scattering of the 
neutrino distribution function (${\cal F}_\nu$) for an initial Gaussian 
distribution centered on 40 MeV, for the thermodynamic characteristics 
specified by StarB in Table \ref{tab:star}.  The curves show the 
distribution at snapshots in time: (A) $t\,=\,0.33\,\mu$s, 
(B) $t\,=\,1.00\,\mu$s, (C) $t\,=\,3.30\,\mu$s, 
(D) $t\,=\,10.0\,\mu$s, (E) $t\,=\,33.0\,\mu$s, and (F) $t\,=\,1000\,\mu$s. 
The solid dots denote an equilibrium Fermi-Dirac 
distribution at the temperature of the surrounding thermal bath with a 
neutrino chemical potential $\mu_\nu\simeq2.32T$ set by the initial 
$\nu_\mu$ neutrino number density.\label{fig:ns_star36_fnu}}
\end{figure}

\begin{figure} 
\vspace*{7.50in}
\hbox to\hsize{\hfill\includegraphics{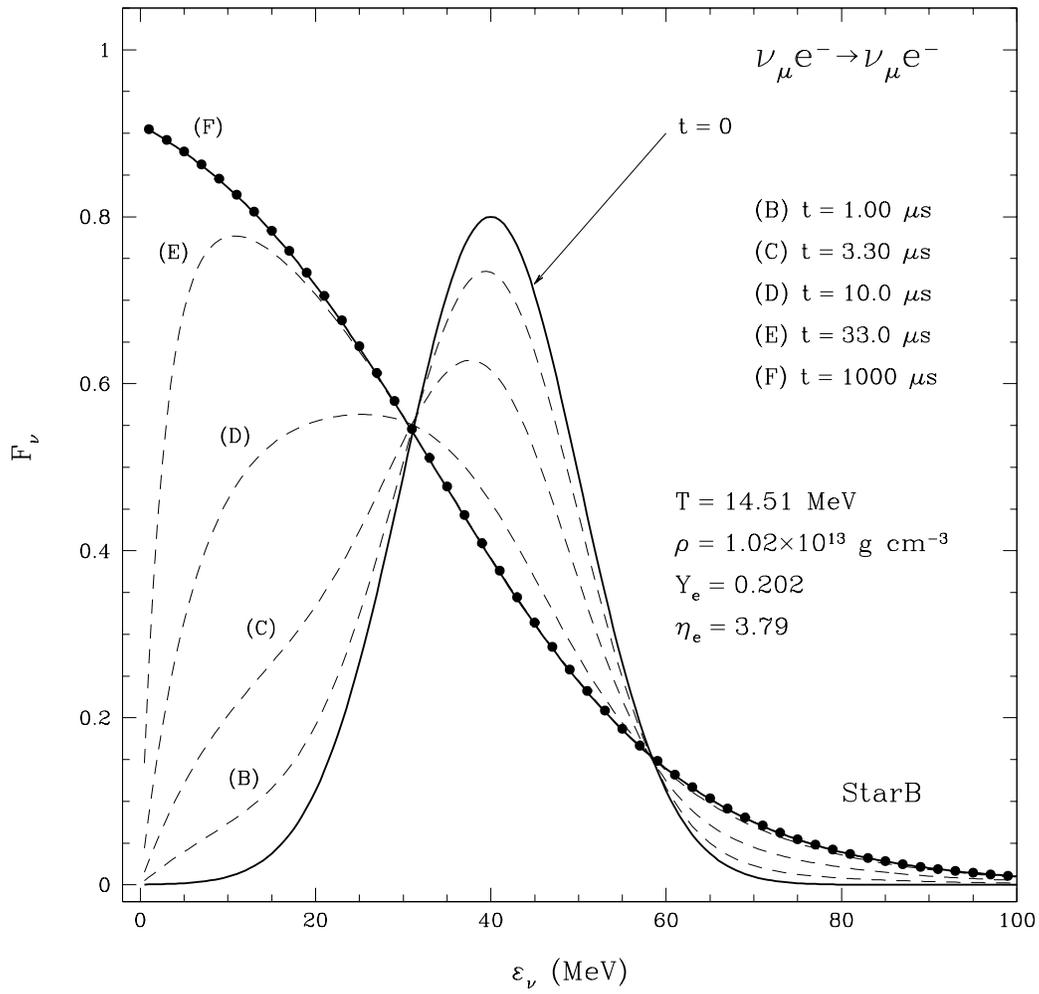}\kern+6in\hfill}
\caption{The same as Fig. \ref{fig:ns_star36_fnu}, but for 
$\nu_\mu$-electron scattering.\label{fig:es_star36_fnu}}
\end{figure}

\begin{figure} 
\vspace*{7.50in}
\hbox to\hsize{\hfill\includegraphics{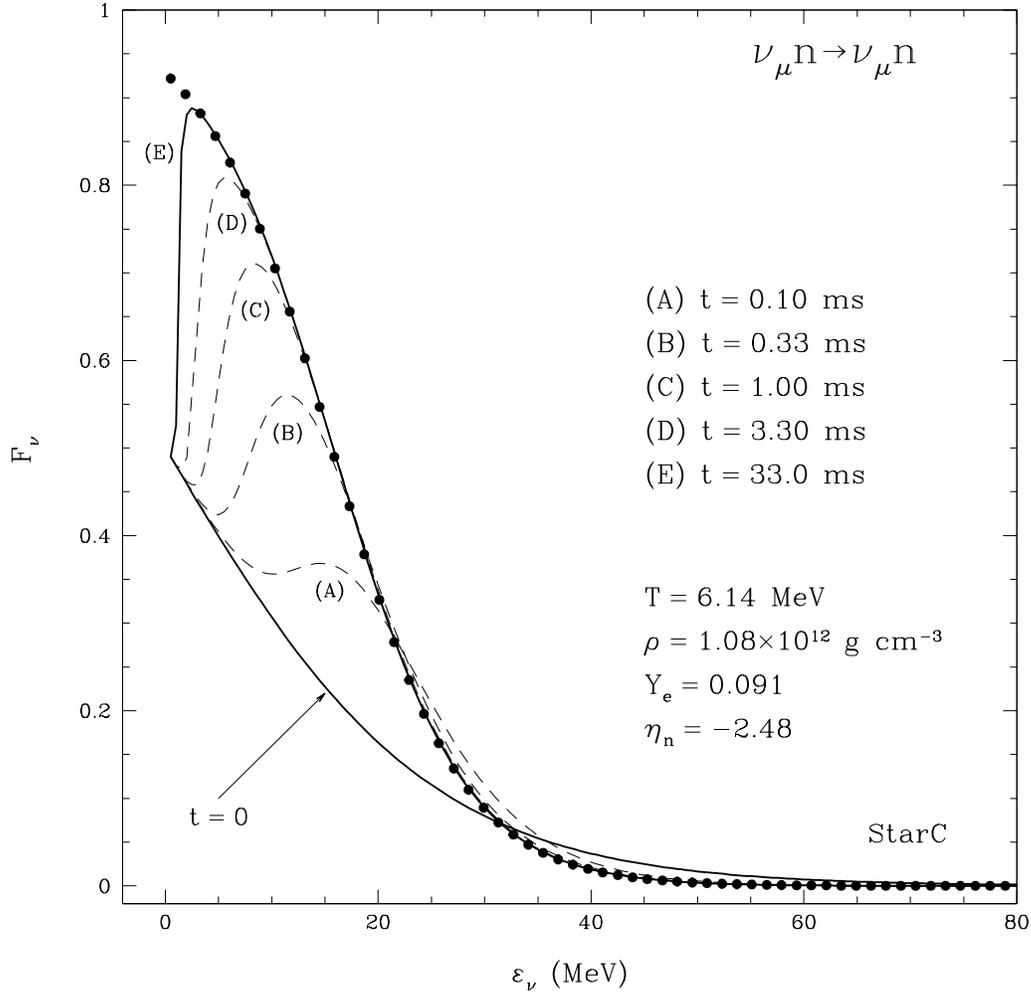}\kern+6in\hfill}
\caption{The time evolution via $\nu_\mu$-neutron scattering of the 
neutrino distribution function (${\cal F}_\nu$) for 
an initial Fermi-Dirac distribution at 2$\times$ the ambient 
temperature, for the thermodynamic characteristics specified by 
StarC in Table \ref{tab:star}.  The curves show the distribution 
at snapshots in time: (A) $t\,=\,0.10$ milliseconds (ms), 
(B) $t\,=\,0.33$ ms, (C) $t\,=\,1.0$ ms, 
(D) $t\,=\,3.3$ ms, and (E) $t\,=\,33.0$ ms. The solid dots 
denote an equilibrium Fermi-Dirac distribution at the temperature 
of the surrounding thermal bath with a neutrino chemical potential 
$\mu_\nu\simeq2.55T$ set by the initial $\nu_\mu$ neutrino number 
density.  Comparison of this plot with 
Fig. \ref{fig:es_star68_newfd200_fnu} shows that $\nu_\mu n$ 
scattering dominates thermalization 
above $\varepsilon_\nu\sim10$ MeV.}
\label{fig:ns_star68_newfd200_fnu}
\end{figure}

\begin{figure} 
\vspace*{7.50in}
\hbox to\hsize{\hfill\includegraphics{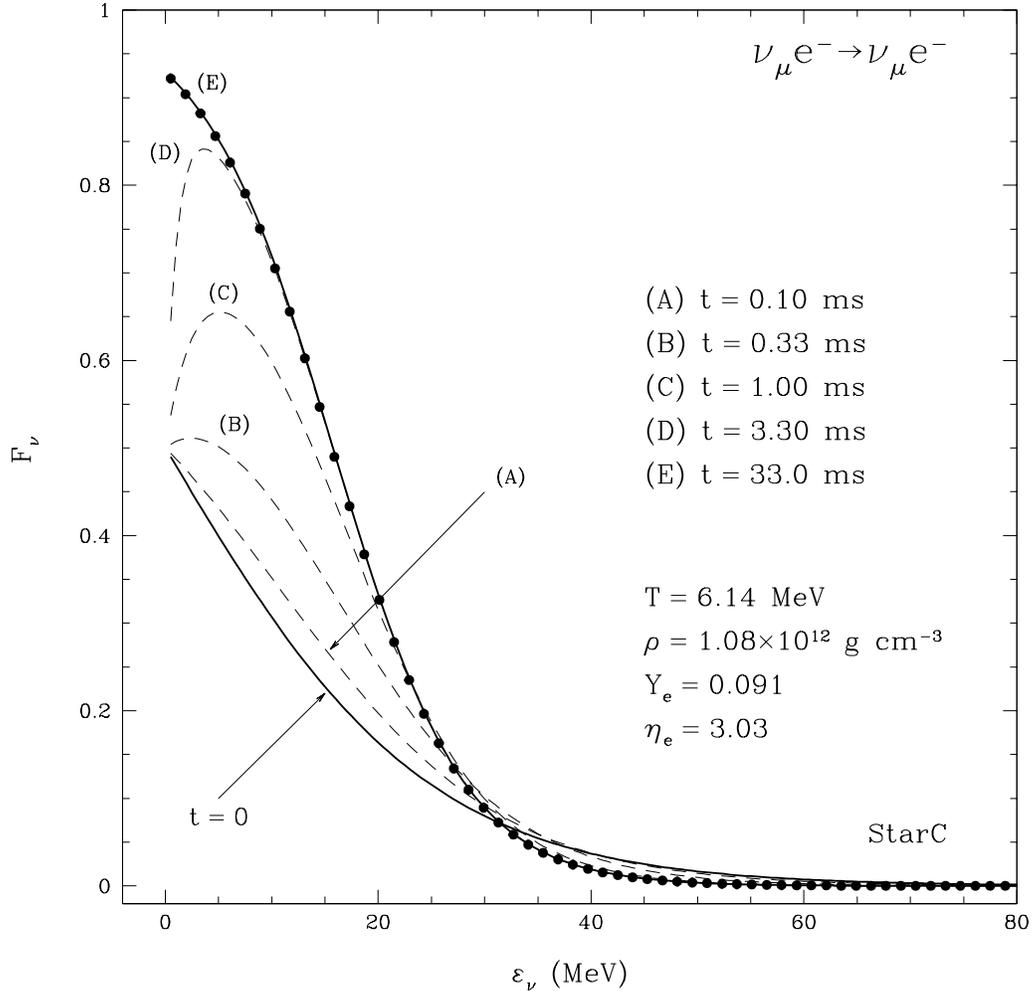}\kern+6in\hfill}
\caption{The same as Fig. \ref{fig:ns_star68_newfd200_fnu}, 
but for $\nu_\mu$-electron scattering.  Comparison of this plot with Fig. 
\ref{fig:ns_star68_newfd200_fnu} shows that $\nu_\mu$-electron 
scattering dominates thermalization below $\varepsilon_\nu\sim 10$ MeV.}
\label{fig:es_star68_newfd200_fnu}
\end{figure}

\begin{figure} 
\vspace*{7.50in}
\hbox to\hsize{\hfill\includegraphics{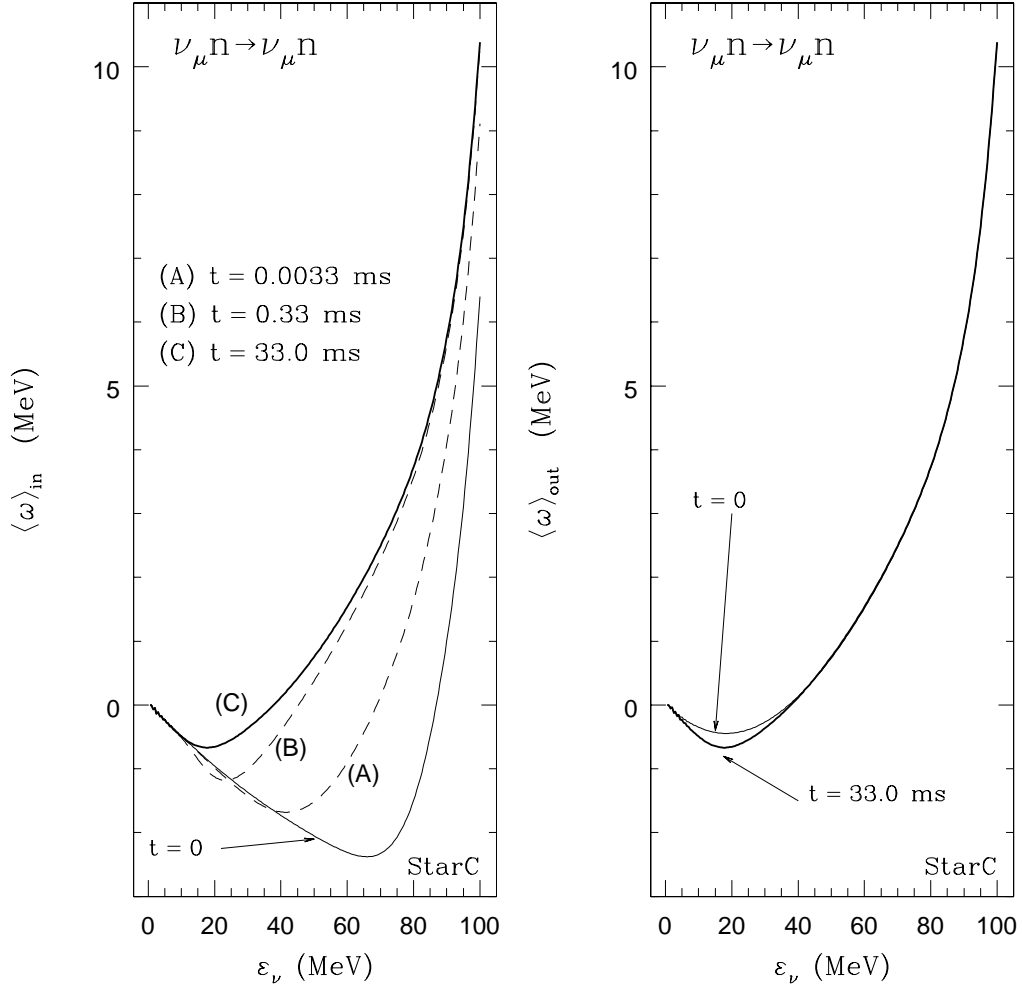}\kern+6in\hfill}
\caption{The thermal average energy transfers, $\langle\omega\rangle_{in}$ 
and $\langle\omega\rangle_{out}$, defined in 
eqs. (\ref{win}) and (\ref{wout}), 
respectively, as a function of neutrino energy ($\varepsilon_\nu$) for 
$\nu_\mu$-neutron scattering at the thermodynamic point StarC.  
The curves show snapshots of the average energy transfers in
time as ${\cal F}_\nu$ evolves (see Fig. \ref{fig:ns_star68_newfd200_fnu}). 
For $\langle\omega\rangle_{in}$, (A) $t\,=\,0.0033$ milliseconds (ms), 
(B) $t\,=\,0.33$ ms, and (C) $t\,=\,33.0$ ms.  
We show $\langle\omega\rangle_{out}$ at $t=0$ (thin line) and 
$t=33.0$ ms (thick line).  Note that in equilibrium ($t\sim33.0$ ms)
$\langle\omega\rangle_{in}=\langle\omega\rangle_{out}$.}
\label{fig:ns_star68_newfd200_winout} 
\end{figure}

\begin{figure} 
\vspace*{7.50in}
\hbox to\hsize{\hfill\includegraphics{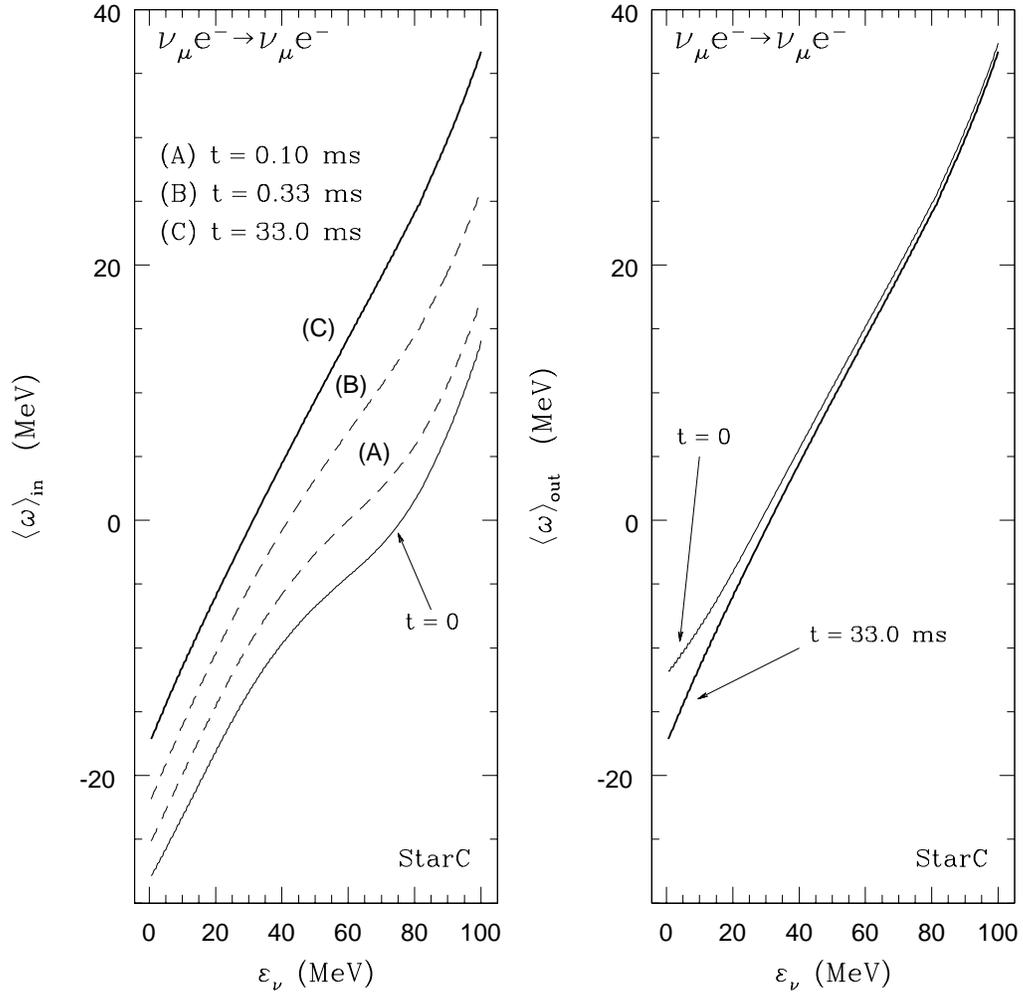}\kern+6in\hfill}
\caption{
The same as Fig. \ref{fig:ns_star68_newfd200_winout}, but for 
$\nu_\mu$-electron scattering.  For $\langle\omega\rangle_{in}$, 
(A) $t\,=\,0.10$ milliseconds (ms), (B) $t\,=\,0.33$ ms, 
and (C) $t\,=\,33.0$ ms.}
\label{fig:es_star68_newfd200_winout} 
\end{figure}

\begin{figure} 
\vspace*{7.0in}
\hbox to\hsize{\hfill\includegraphics{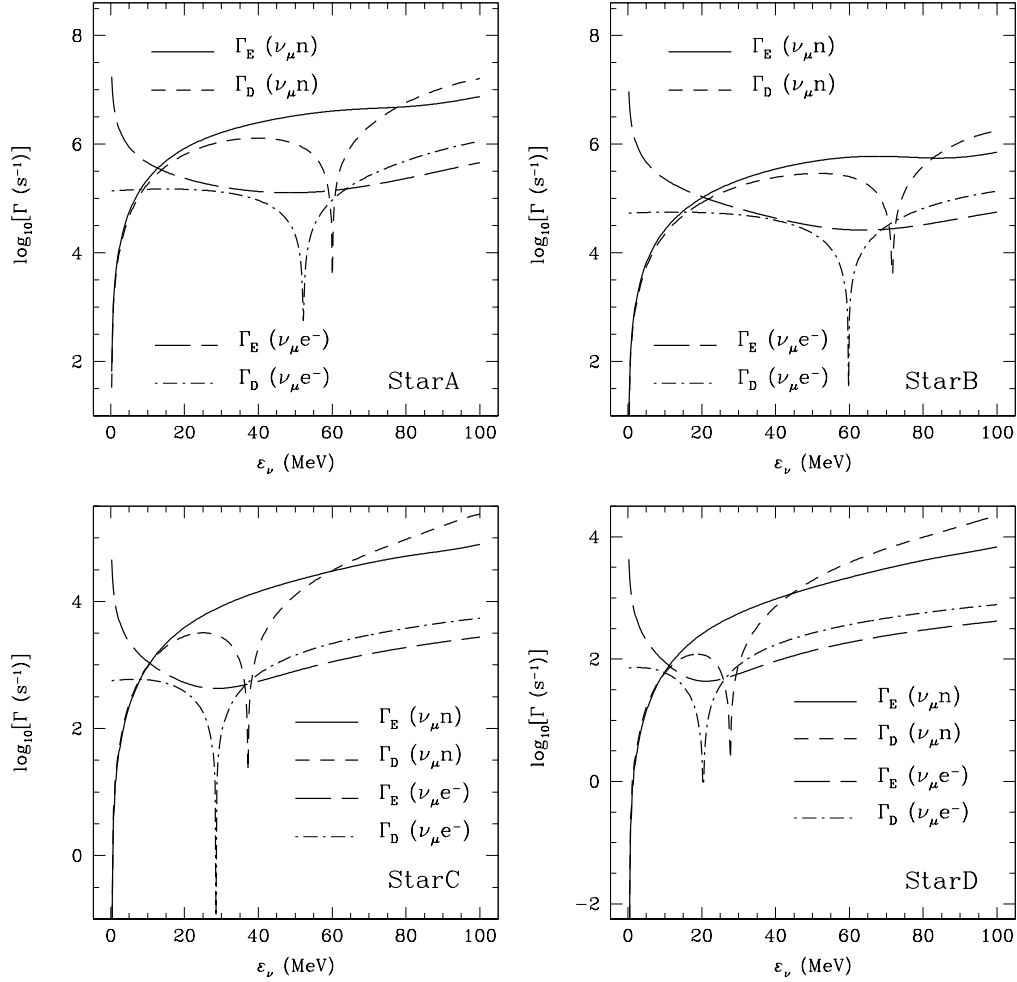}\kern+6in\hfill}
\caption{$\Gamma_D$ and $\Gamma_E$ as defined in eqs. (\ref{gammad}) 
and (\ref{gammae}), respectively, for both $\nu_\mu n$ and $\nu_\mu e^-$ 
scattering for an initial Fermi-Dirac distribution at StarA, StarB, StarC, 
and StarD (see Table \ref{tab:star}) at a snapshot in time.  
The spikes in the $\Gamma_D$ curves are a consequence of the fact that 
$\langle\omega\rangle_{out}\rightarrow0$ at those neutrino energies 
(compare the plot above for StarA with 
Figs. \ref{fig:ns_star68_newfd200_winout} and 
\ref{fig:es_star68_newfd200_winout}).  
The solid and the short-dashed lines in all four plots are 
$\Gamma_E$ and $\Gamma_D$, respectively, for $\nu_\mu n$ scattering.  
The long-dashed and long-short-dashed lines are $\Gamma_E$ and $\Gamma_D$,
respectively, for $\nu_\mu e^-$ scattering.  At all four points in the 
stellar profile {\it Star}, $\nu_\mu n$ scattering
dominates $\nu_\mu e^-$ scattering at energies above 10$-$20 MeV by 
approximately an order of magnitude.  For StarA, the points where the 
rates for $\nu_\mu$-electron and $\nu_\mu$-neutron scattering cross 
are at $\sim8$ and $\sim13$ MeV.  For StarB, they lie higher, at 
$\sim15$ and $\sim20$ MeV, due predominantly to the higher temperature 
at that radius.  For both StarC and StarD, 
the rates cross at between $7$ and $12$ MeV.  }
\label{fig:nses_allstar_newfd400_tub}
\end{figure}

\begin{figure} 
\vspace*{7.50in}
\hbox to\hsize{\hfill\includegraphics{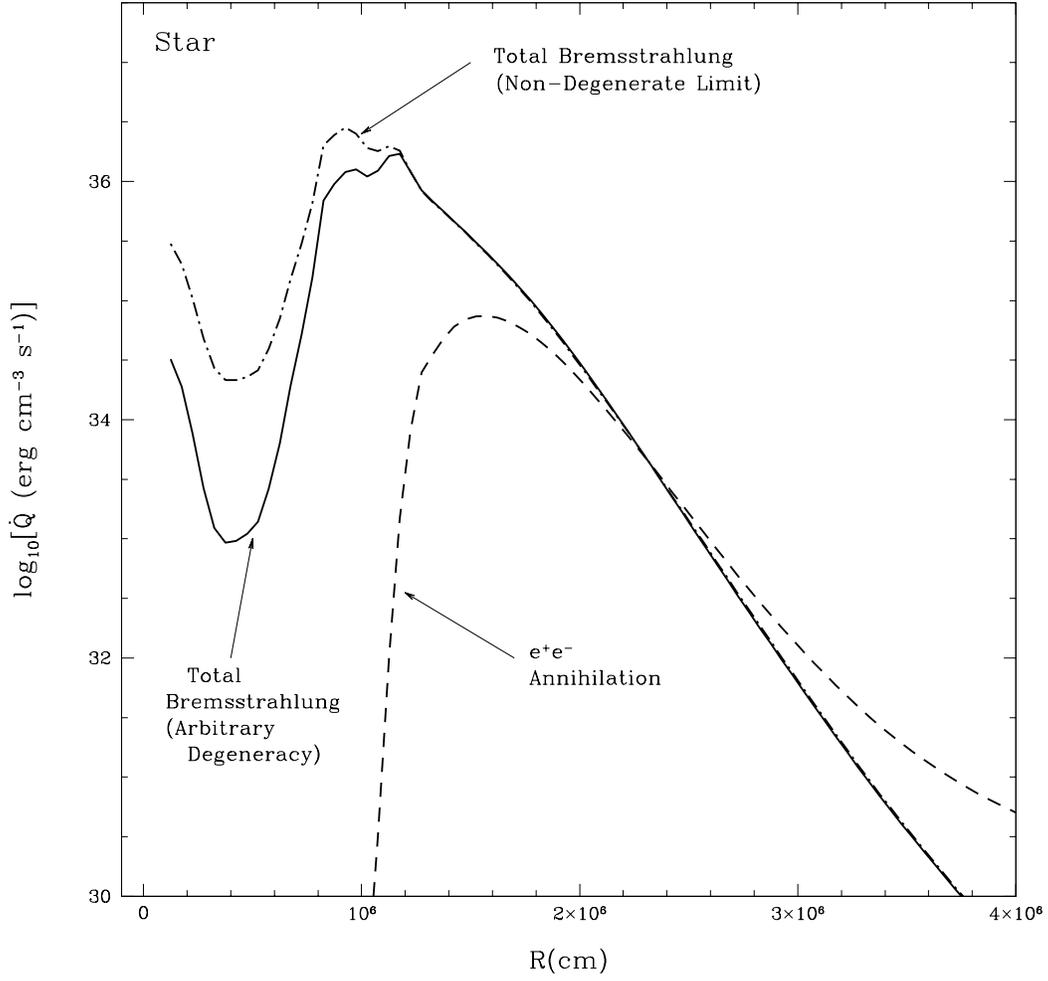}\kern+6in\hfill}
\caption{The integrated total volumetric emissivity in 
ergs cm$^{-3}$ s$^{-1}$ for $e^+e^-$ annihilation and
nucleon-nucleon bremsstrahlung in the non-degenerate limit 
(Eq. \ref{totalbremspec}) and at arbitrary degeneracy 
(Eq. \ref{totbrem2}) as a function of radius in the stellar collapse 
profile, {\it Star}.  Note that the total bremsstrahlung
and $e^+e^-$ annihilation emissivities cross at $R\sim 23$ kilometers 
where $\rho\simeq6\times10^{12}$ g cm$^{-3}$ and
$T\simeq11$ MeV.  Above this radius, $e^+e^-$ dominates.}
\label{fig:radcomp_star}
\end{figure}

\begin{figure} 
\vspace*{7.50in}
\hbox to\hsize{\hfill\includegraphics{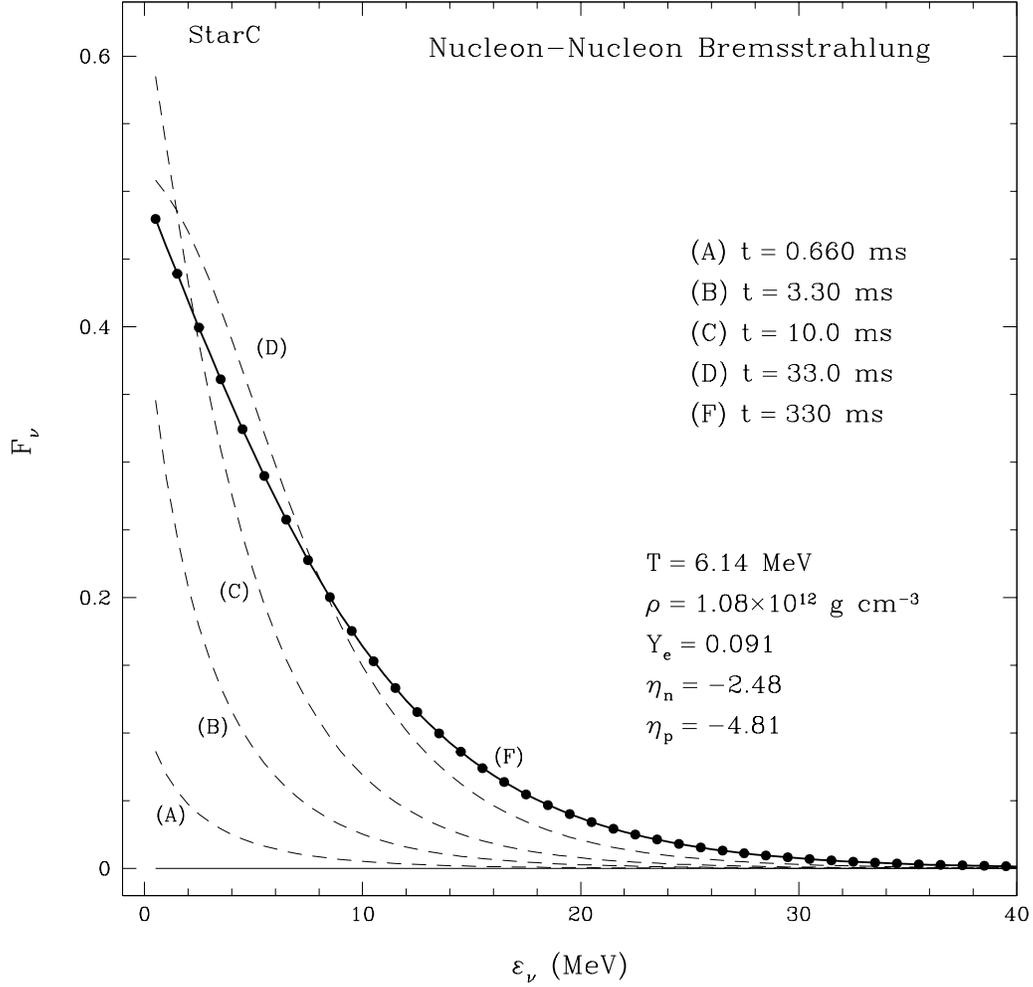}\kern+6in\hfill}
\caption{The time evolution of ${\cal F}_\nu$ due to nucleon-nucleon 
bremsstrahlung via eq. (\ref{workingbrem}) for
the point StarC described in Table \ref{tab:star} starting 
with ${\cal F}_\nu={\cal F}_{\bar{\nu}}=0$ at all energies. Curves 
show snapshots of the evolution of ${\cal F}_\nu$ with time: 
(A) $t\,=\,0.66$ milliseconds (ms), (B) $t\,=\,3.30$ ms, 
(C) $t\,=\,10.0$ ms, (D) $t\,=\,33.0$ ms, and (F) $t\,=\,330$ ms. 
The solid dots denote an equilibrium Fermi-Dirac distribution at 
the temperature of the surrounding thermal bath with  zero 
neutrino chemical potential.}
\label{fig:br_star68_fnu}
\end{figure}

\begin{figure} 
\vspace*{7.50in}
\hbox to\hsize{\hfill\includegraphics{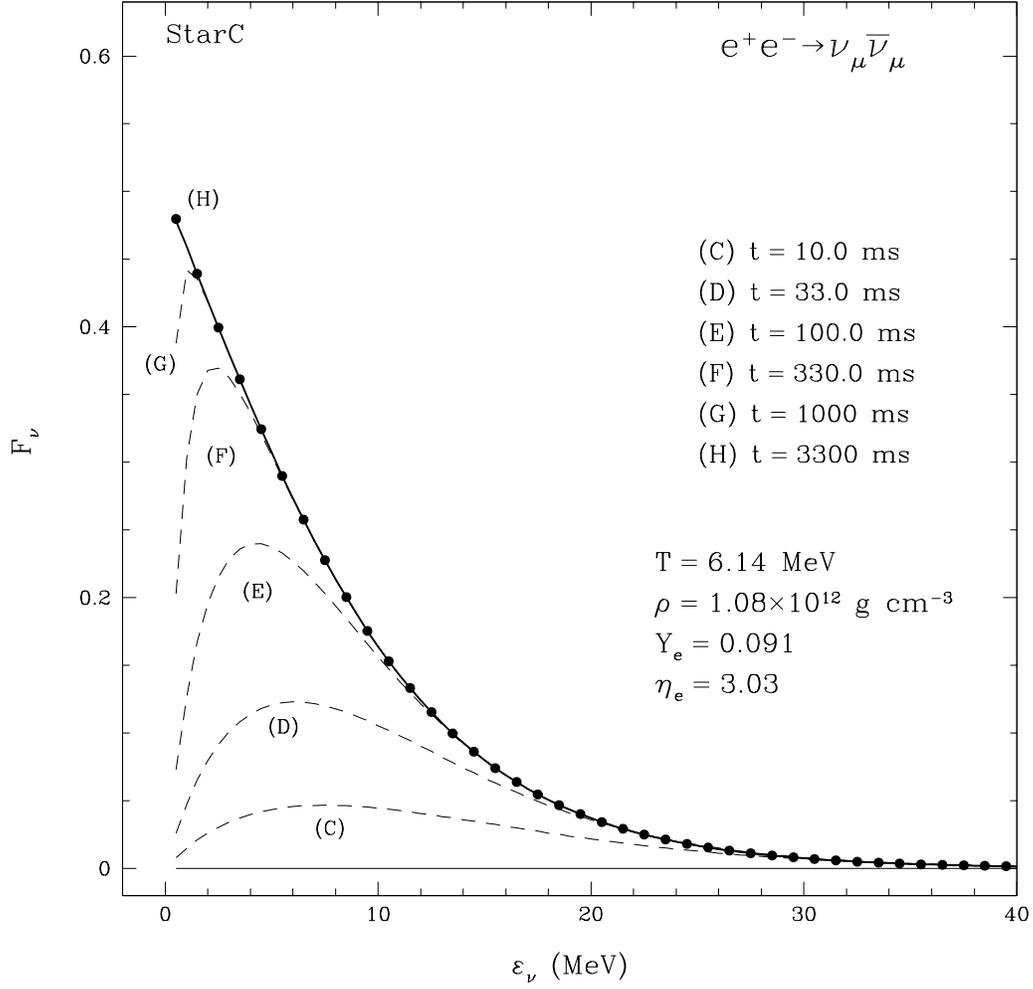}\kern+6in\hfill}
\caption{The same as Fig. \ref{fig:br_star68_fnu}, but for 
$e^+e^-$ annihilation via eq. (\ref{workingpairpr}).
Curves show snapshots of the evolution of ${\cal F}_\nu$ with time: 
(C) $t\,=\,10.0$ milliseconds (ms), (D) $t\,=\,33.0$ ms, 
(E) $t\,=\,100.0$ ms, (F) $t\,=\,330.0$ ms, (G) $t\,=\,1000$ ms, 
and (H) $t\,=\,3300$ ms.}
\label{fig:pp_star68_fnu}
\end{figure}

\begin{figure} 
\vspace*{7.50in}
\hbox to\hsize{\hfill\includegraphics{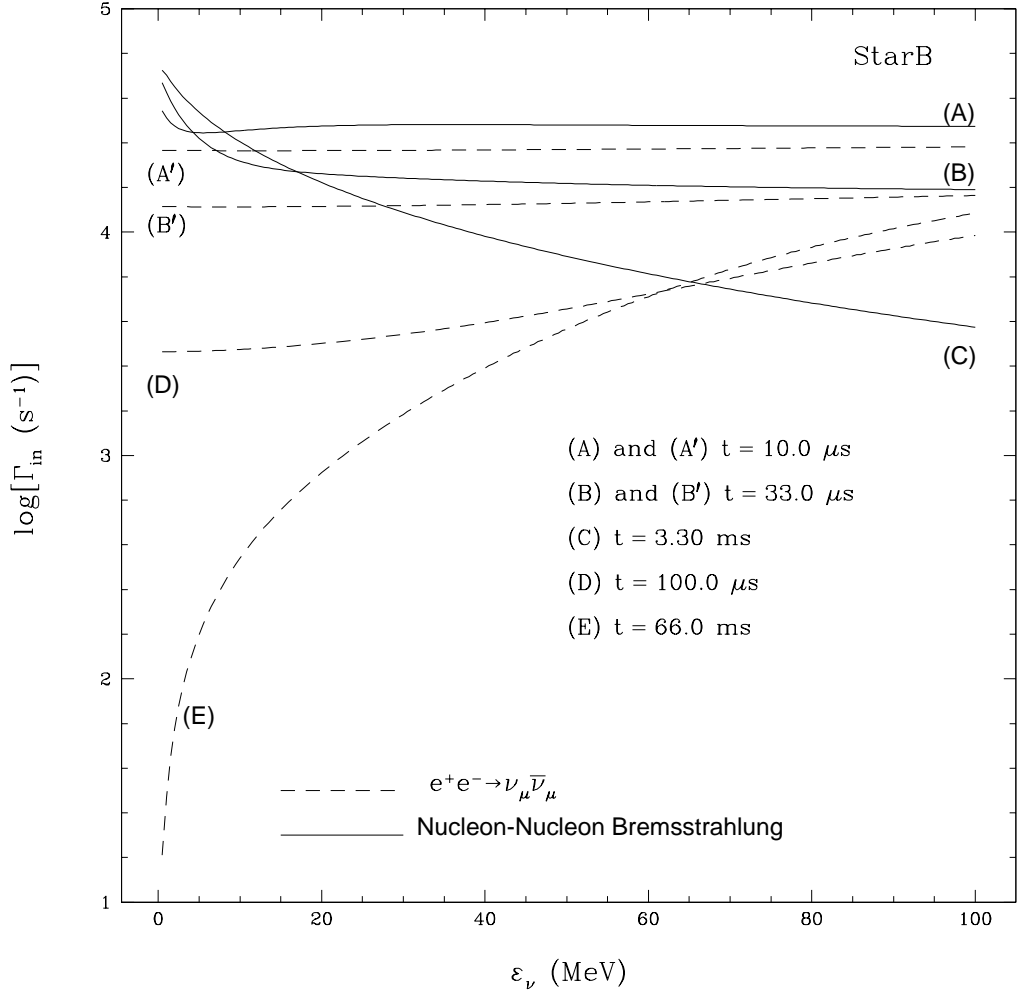}\kern+6in\hfill}
\caption{$\Gamma_{in}$ as defined in eq. (\ref{gammain}) for 
nucleon-nucleon bremsstrahlung (solid lines) and 
$e^+e^-$ annihilation (dashed lines), for the point StarB.  
Each curve shows a snapshot of $\Gamma_{in}$ as ${\cal F}_\nu$ builds
from zero phase-space occupancy at $t=0$: for (A) and (A$\pr$) 
$t=10.0$ $\mu$s. (B) and (B$\pr$) denote $t=33.0$ $\mu$s. (C) marks
the equilibrium rate for bremsstrahlung at $t=3.3$ milliseconds.  
Curves (D) and (E) mark $100.0$ $\mu$s and $66.0$ milliseconds,
respectively, for $e^+e^-$ annihilation.  The latter, marks the 
$e^+e^-$ equilibrium rate.  Note that the equilibrium rates
cross at $\varepsilon_\nu\sim65$ MeV.}
\label{fig:brpp_star36_tin}
\end{figure}

\begin{figure} 
\vspace*{7.50in}
\hbox to\hsize{\hfill\includegraphics{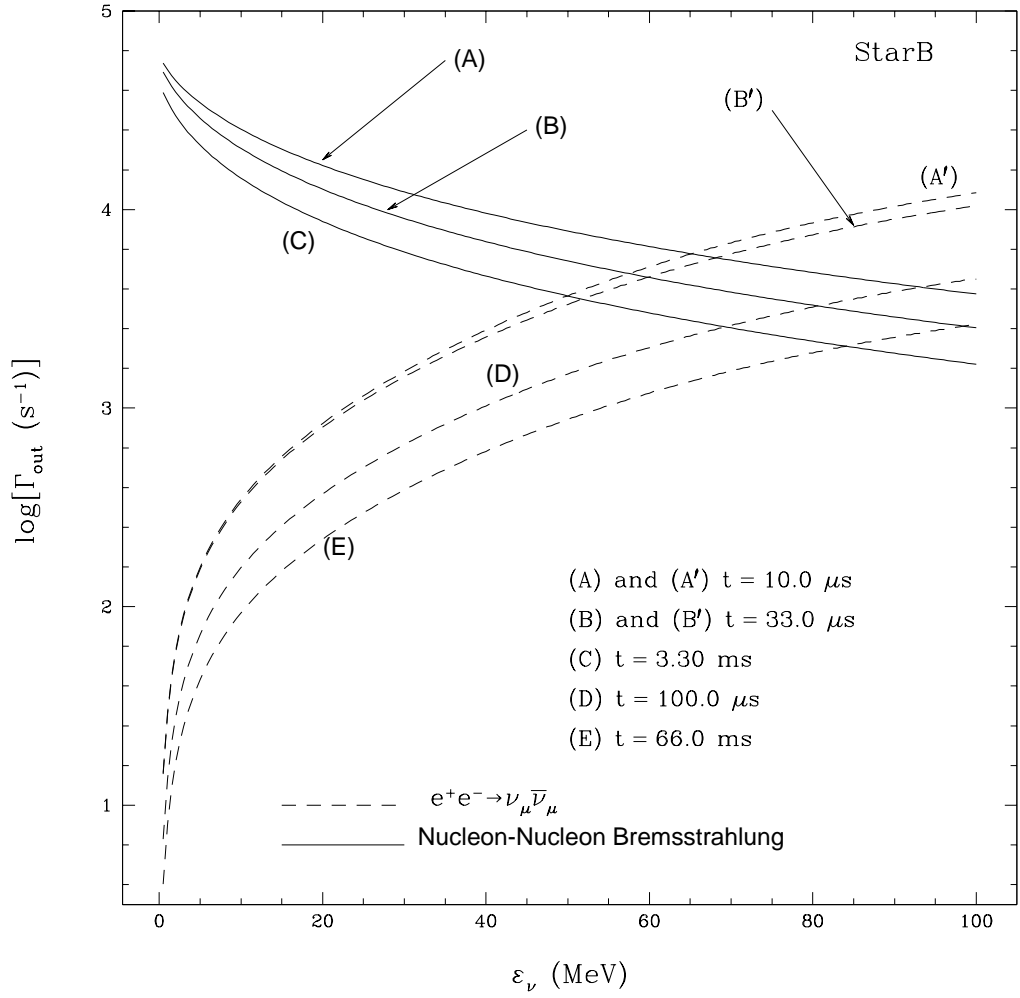}\kern+6in\hfill}
\caption{The same as Fig. \ref{fig:brpp_star36_tin}, for the 
same times, but for $\Gamma_{out}$ as defined in eq. (\ref{gammaout}).}
\label{fig:brpp_star36_tout}
\end{figure}
\end{document}